\documentstyle[12pt,epsf]{article}
\newcommand{\be}{\begin{equation}}
\newcommand{\ee}{\end{equation}}
\newcommand{\bea}{\begin{eqnarray}}
\newcommand{\eea}{\end{eqnarray}}

\newcommand{\ov}{\overline}
\newcommand{\bvec}{\mathbf}
\newcommand{\rt}{\rightarrow}

\newcommand{\slp}{/\!\!\!p}

\newcommand{\gapproxeq}
{\lower .7ex\hbox{$\;\stackrel{\textstyle >}{\sim}\;$}}
\newcommand{\lapproxeq}
{\lower .7ex\hbox{$\;\stackrel{\textstyle <}{\sim}\;$}}

\newcommand{\p}{{\mathrm p}}

\newcommand{\mj}{M_j}

\newcommand{\hh}{\hat{h}}
\newcommand{\hH}{\widehat{H}}
\newcommand{\hmu}{\widehat{M}_u}
\newcommand{\hdmj}{\Delta \widehat{M}_j}
\newcommand{\hgi}{\widehat{\Gamma}_i}
\newcommand{\hgj}{\widehat{\Gamma}_j}
\newcommand{\hrho}{\hat{\rho}}
\newcommand{\hp}{\hat{\mathrm p}}

\def\neb{\hbox{$\ov{\nu}_e \!$ }}

\def\ca{{C_{\scriptscriptstyle A}}}
\def\cv{{C_{\scriptscriptstyle V}}}

\newcommand{\tre}{\left( \cv^2 + 3 \ca^2 \right)}

\newcommand{\Nature}{{\it Nature \,}}

\newcommand{\ApJ}{{\it Astrophys. J. \,}}
\newcommand{\ApJS}{{\it Astrophys. J. Suppl. \,}}

\newcommand{\NP}{{\it Nucl. Phys. \,}}
\newcommand{\PR}{{\it Phys. Rev. \,}}
\newcommand{\PRL}{{\it Phys. Rev. Lett. \,}}

\newcommand{\etal}{{\it et al.}}

\begin{document}
\setlength{\unitlength}{1mm}

{\hfill astro-ph/9906232}

{\hfill DSF 18/99}\vspace*{1cm}

\begin{center}
{\Large \bf Big Bang Nucleosynthesis: an accurate determination of light
element yields}
\end{center}

\bigskip\bigskip

\begin{center}
{\bf S. Esposito}, {\bf G. Mangano}, {\bf G. Miele}, and {\bf O. Pisanti},
\end{center}

\vspace{.5cm}

\noindent
{\it Dipartimento di Fisica, Universit\'{a} di Napoli "Federico II", and INFN,
Sezione di Napoli, Mostra D'Oltremare Pad. 20, I-80125 Napoli, Italy}\\
\bigskip\bigskip\bigskip

\begin{abstract}
We report the results of a new accurate evaluation of light nuclei yields
in primordial nucleosynthesis. All radiative effects, finite nucleon mass,
thermal and plasma corrections are included in the proton to neutron
conversion rates. The relic densities of $^4He$, $D$ and $^7Li$ have been
numerically obtained {\it via} a new updated version of the standard BBN
code. In particular the theoretical uncertainty on $^4He$ is reduced to the
order of 0.1 \%.
\end{abstract}
\vspace*{2cm}

\begin{center}
{\it PACS number(s): 98.80.Cq; 95.30.Cq; 11.10.Wx; 13.40.Ks}
\end{center}

\newpage
\baselineskip=.8cm

\section{Introduction}
\setcounter{equation}0

Big Bang Nucleosynthesis (BBN) still represents one of the key subject of
modern cosmology even if its clear understanding traces back to over 25
years ago \cite{Wagoner}. The reason for this relies on the fact that BBN
is one of the most powerful tools to study fundamental interactions, since
light nuclei abundances are crucially depending on many elementary particle
properties. As a well-known example, the $^4He$ abundance is strongly
affected by the number of {\it effective} neutrino degrees of freedom, but
others fascinating phenomena such as neutrino degeneracy or oscillation
phenomena can be studied too, using the universe few seconds after the bang
as a laboratory. 

In the recent years, the experimental accuracy of the
light primordial nuclei abundances, mainly the one of $^4He$, underwent a
sort of revolution. The qualitative results of not too many years ago,
suggesting that the $^4He$ mass fraction $Y_4$ was of the order of 0.25,
recently turned  in measurements with accuracies of the order
of one percent. A similar good improvement has been obtained in both
Deuterium ($D$) and $^7Li$ abundances $Y_2 \equiv D/H$ and $Y_7 \equiv
~^7Li/H$. In particular for $D$, measurements in distant Quasars Absorption
line Systems (QAS) now represent a reliable estimate of the primordial
value for $Y_2$, which is only lowered by subsequent stellar processing.
Paradoxically, the refinement of these experimental techniques, due to
the uncertainties in the models describing stellar
activity, is at the basis of large discrepancies between different set of
results. Such discrepancies are possibly of systematic origin, or may
reveal new aspects of cosmological evolution of the universe. The
observations of $Y_4$ from regression to zero metallicity in blue compact
galaxies in two independent surveys still produce two incompatible
results, a {\it low } value \cite{Steigman},
\be
Y_4^{(l)} = 0.234 {\pm} 0.002 {\pm} 0.005~~~,
\label{e:lowHe}
\ee
and a significantly higher one \cite{Izotov},
\be
Y_4^{(h)} = 0.243 {\pm} 0.003~~~.
\label{e:highHe}
\ee
A similar situation occurs in $D$ measurements, where observations in
different QAS, both at red shift larger than 3, give two results at bias
for one order of magnitude \cite{Songaila,Tytler}
\bea
Y_2^{(l)} &=& \left(3.4 {\pm} 0.3 \right) 10^{-5}~~~,
\label{e:lowD} \\
Y_2^{(h)} &=& \left(1.9 {\pm} 0.4 \right) 10^{-4}~~~.
\label{e:highD}
\eea
For the $^7Li$ abundance, the almost constant {\it Spite
plateau} observed in the halo of POP II stars \cite{Thorburn, Molaro}
\be
Y_7^{(l)} =\left(1.6 {\pm} 0.36 \right) 10^{-10}~~~,
\label{e:lowLi7}
\ee
is generally considered a reliable estimate of primordial abundance.
Nevertheless, the observation of stars similar to the ones contributing
to the Spite plateau, but with no traces of $^7Li$ \cite{Thorburn, Ryan},
seems to imply the presence of a depletion mechanism. A recent analysis
based on a sample of 41 stars does not find any evidence of depletion
mechanism or post-BBN creation and yields the primordial abundance
\cite{Bonifacio}
\be
Y_7^{(h)} =\left(1.73 {\pm} 0.21 \right) 10^{-10}~~~.
\label{e:higLi7}
\ee
A brief summary of the complete experimental situation on primordial
abundances can be found in Ref. \cite{Sarkar}.

Probably, future measurements or a better understanding of the present data
will clarify about the systematics. Nevertheless what is emerging from the
above results is that the $^4He$ data are reaching a precision of the order
of few percents. This fact requires a similar effort in the theoretical
analysis, in order to reduce the uncertainty on the predictions at least at 
the same level of magnitude. In a previous paper \cite{EMMP1} we 
performed a thoroughly analysis of all corrections to the proton/neutron
conversion rates,
\bea
(a)~~~ \nu_e + n \rightarrow e^- + p &~~~,~~~~~~& (d)~~~ \neb + p
\rightarrow e^+ + n~~~, \nonumber \\
(b)~~~e^- + p \rightarrow \nu_e + n &~~~,~~~~~~& (e) ~~~n \rightarrow e^- +
\neb + p ~~~,\nonumber \\
(c)~~~ e^+ + n \rightarrow \neb + p &~~~,~~~~~~& (f)~~~ e^- + \neb + p
\rightarrow n~~~,
\label{e:reaction}
\eea
which fix at the freeze out temperature $\sim 1~MeV$ the neutron to proton
density ratio. The Born rates, obtained in the tree level $V-A$ limit and
with infinite nucleon mass, have been corrected to take into account
basically three classes of relevant effects:
\begin{itemize}
\item[{i)}] electromagnetic radiative corrections,
which largely contribute to the rates of
the fundamental processes, in particular in the low temperature regime,
$T\leq 0.1 ~MeV$;
\item[{ii)}] finite nucleon mass corrections, which are of the order of
$T/M_N$ or $m_e/M_N$, with $m_e,M_N$ the electron and nucleon mass,
respectively;
\item[{iii)}] plasma effects, proportional to the surrounding plasma
temperature, which both affect the microscopic process rates $(a)-(f)$, as
well as the neutrino to photon temperature ratio through $e^{\pm}$,
$\gamma$ equations of state.
\end{itemize}
The other main source of theoretical uncertainty comes from the partial
knowledge of nuclear rates relevant for nuclei formation. Their numerical
expressions, obtained by a convolution of the experimental data with a
Boltzmann distribution, are affected by uncertainties of the order of
$10\%$ (see references quoted in
\cite{Kawano}). More crucially, in many cases, these fits are known
to well describe the data in a temperature interval which is only partially
overlapping the one relevant for BBN, $0.01 ~MeV \leq T \leq 10 ~MeV$.
However, both a Montecarlo analysis to sample the error distribution of the
reaction cross sections \cite{Montecarlo}, and a more recent method based
on linear error propagation \cite{Fiorentini}, show that, in particular for
$^4He$ mass fraction, the effect is {at most} as large as the one due to
the uncertainty on neutron lifetime $\tau_n$, and smaller than $1\%$.
Therefore it is theoretically justified to look, as in \cite{EMMP1}, for
all sources of theoretical uncertainty up to this level of precision. The
situation gets worse with $D$ and $^7Li$, where the uncertainties due
nuclear reactions can be as large as $(10 \div 30)\%$ \cite{Fiorentini}.

This paper represents the natural companion to \cite{EMMP1}. We have
built a new updated version of the standard BBN code, which is available
since many years \cite{Wagoner,Kawano}, where all corrections i)-iii) have
been included. In particular we have also included the modified $e^{\pm}$, 
$\gamma$ equations of state
due to electromagnetic mass renormalization. In Section 2 we
review the corrections to $n\leftrightarrow p$ Born rates, while in Section
4 we discuss the numerical method we have used to integrate the set of
equations relevant for BBN, which are described in Section 3. The numerical
results for light nuclei abundances, as functions of the final baryon to
photon density ratio, $\eta$, the number of effective neutrino degrees of
freedom, $N_\nu$, and the neutron lifetime, $\tau_n$, are reported in Section
5, where they are discussed and compared with the experimental data. We
have also performed a fit of these abundances with a precision of the order
of $0.1\%$ in the interesting range for the parameters $\eta$, $N_\nu$ and
$\tau_n$. Finally in Section 6 we give our conclusions.

\section{Corrections to $n \leftrightarrow p$ Born rates}
\setcounter{equation}0

As is well known, the key parameter in determining the primordial $^4He$
mass fraction, $Y_4$, is the value of the neutron to proton density ratio
at the freeze-out temperature $T \sim 1 ~MeV$, since almost all residual
neutrons are captured in $^4He$ nuclei due to its large binding energy per
nucleon. In order to make an accurate theoretical prediction for $Y_4$ it
is necessary, though not sufficient, to have a reliable evaluation of the
rates for the processes (\ref{e:reaction}). An effort in this direction has
been pursued in the last ten years by many authors. Recently, the entire
set of corrections to the Born rates $\omega_B$ at the level of $1 \%$
accuracy have been recalculated in \cite{EMMP1} and \cite{Lopez}, with
quite compatible results. In this Section we shortly summarize the main
corrections $\Delta \omega /\omega_B$ coming from considering radiative,
finite nucleon mass, and thermal effects. This short review is here
included for the sake of completeness and to fix the notation. A detailed
discussion of the subject can be found in our paper \cite{EMMP1}.

\subsection{The Born rates}

Let us consider as an example, the thermal averaged rate per nucleon for
the neutron decay process $(e)$. In the simple $V-A$ tree level, and in the
limit of infinite nucleon mass, which we will refer to as {\it Born
approximation}, one has
\be
\omega_B ( n \rightarrow e^- + \neb + p ) \, = \,  \frac{G_F^2 \tre}{2
\pi^3} \, \int_0^\infty d {|\bvec{p}'| \,|\bvec{p}'|^2} \,  q_0^2 \,
\Theta(q_0) \,\left[ 1 - F_\nu (q_0)\right]  \left[ 1 - F_e(p_0') \right],
\label{e:cb15}
\ee
where $G_F$ is the Fermi coupling constant, $\cv$ and $\ca$ the nucleon
vector and axial coupling. In our notation $\bvec{p}'$ and $p_0'$
are the electron momentum and energy, and $q_0$ the neutrino energy. The
integration limits are imposed by the $\Theta$-function, $q_0 \geq 0$. For
reaction $(e)$ we have $q_0 = M_n - M_p - p_0' \equiv \Delta - p_0'$. The
Fermi statistical distributions for $e^{\pm}$ and neutrinos in the {\it
comoving frame}, neglecting chemical potentials, are
\be
F_e(p_0') = \left[e^{\beta |p_0'|} +1 \right]^{-1}~~~,~~~~~~~~~ F_\nu(q_0)
= \left[e^{\beta_\nu |q_0|} +1 \right]^{-1}~~~,
\label{e:cbb}
\ee
with $\beta=1/T$ and $\beta_\nu = 1/T_\nu$ \footnote{The ratio $T_\nu / T$
is fixed by entropy conservation and using the neutrino decoupling
temperature \cite{EMMP1} (see Section \ref{s:nut}).}. All other rates for
processes $(a)-(d), (f)$ can be simply obtained from (\ref{e:cb15})
properly changing the statistical factors and the expression for $q_0$
\cite{EMMP1}.

The accuracy of Born approximation can be tested by comparing, for example,
the prediction for neutron lifetime with the experimental value
$\tau_n^{ex} = (886.7 {\pm} 1.9)~s$ \cite{PDG98}. Using $\cv = 0.9751 {\pm} 0.0006$
and $\ca/\cv = 1.2601 {\pm} 0.0025$ \cite{PDG98}, Eq. (\ref{e:cb15}) in the
vanishing density limit gives $\tau_n \simeq 961~s$. Therefore, to recover
the experimental value, a correction of about $8\%$ is expected to come
from radiative and/or finite nucleon mass effects. In the same way these
corrections are also expected to contribute to all six processes $(a)$-$(f)$
relevant for BBN. In addition to these, microscopic $n \leftrightarrow p$
reactions taking place in the early universe, also feel the presence
of the surrounding plasma of $\gamma$ and
$e^{\pm}$ pairs in thermodynamical equilibrium. Emission and absorption of real
$\gamma$ or $e^{\pm}$ from the thermal bath can be taken into account using the
finite temperature field theory in the real time formalism. This has been
considered by several authors \cite{FTQFT}, and recently in \cite{EMMP1}.

\subsection{Electromagnetic radiative corrections}
\label{s:rad}

It is customary to separate the electromagnetic radiative corrections to
the Born amplitudes for processes (\ref{e:reaction}) in $outer$ and $inner$
terms. The first ones involve the nucleon as a whole and consist in a
multiplicative factor to the modulus squared of transition amplitude 
of the form
\be
1+ \frac{\alpha}{2 \pi} g(p_0',q_0)~~~.
\label{e:outer}
\ee
The function $g(p_0',q_0)$
\cite{Sirlin} depends on electron and neutrino energies and describes the
deformation in the electron spectrum. Its effect on a freely decaying
neutron is such to reduce the Born prediction for the lifetime of about
$1.6 \%$.

Inner corrections are sensible to nucleon structure details, and thus much
more difficult to handle. They have been estimated in \cite{Marciano},
studying corrections to the quark weak currents. Translating the
quark--based description in the hadronic language, the inner corrections
result in the additional multiplicative factor
\be
1 + \frac{\alpha}{2 \pi} \left( 4 \, \ln \frac{M_Z}{M_p} \, + \, \ln
\frac{M_p}{M_A} \, + \, 2 C \, + A_g \, \right)~~~,
\label{e:inner}
\ee
where the first term is the short--distance contribution and $A_g=-0.34$ is
a perturbative QCD correction. The other two terms are related to the
axial--induced contributions, with $M_A= (400 \div 1600) MeV$ a low energy
cut-off in the short-distance part of the $\gamma W$ box diagram, and $C$
related to the remaining long distance term.

The global effect of these two kind of corrections, improved by 
resumming all leading logarithmic corrections $\alpha^n ln^n(M_Z)$
\cite{Marciano2}, is {\it via} the multiplicative factor
\be
{\cal G}(p_0',q_0) = \left[ 1 \, + \,  \frac{\alpha}{2 \pi} \left( \ln
\frac{M_p}{M_A} \, + \, 2 C  \right) \, + \, \frac{\alpha (M_p)}{2 \pi} \,
\left[ g(p_0',q_0) \, + \, A_g \right] \right] S(M_p , M_Z)~~~,
\label{e:inout}
\end{equation}
where $\alpha (\mu)$ is the QED running coupling constant defined in the
$\ov{MS}$ scheme and $S(M_p,M_Z)$ a short distance rescaling factor, defined 
in \cite{EMMP1}.

Another effect to be considered, which can be in fact as large as few
percents of the Born rates, is the so-called {\it Coulomb correction}, due
to the rescattering of the electron in the field of the proton and leading to
the Fermi function for Coulomb scattering
\be
{\cal F}(p_0') \, \simeq \, \left( 1 \, + \, \alpha \pi
\, \frac{p_0'}{{|\bvec{p}'|}} \right)~~~.
\label{e:coulomb}
\ee
However, this effect is only present when both electron and proton are in
either the initial or final states, namely it only corrects the amplitudes
of processes $(a), (b), (e)$ and $(f)$.

One may wonder if including the effects given by (\ref{e:inout}) and
(\ref{e:coulomb}) the theoretical prediction for neutron decay is now
compatible with the experimental results. Evaluating numerically the
integral over the phase space one finds $\tau_n^{th} = 893.9~s$, still at
variance with the experiment. Even adding all known sub--leading effects
the agreement does not really improve \cite{Wilkinson}. As in Ref.
\cite{EMMP1} we take the point of view of rescaling all the rates
(\ref{e:reaction}), after including finite nucleon mass corrections (see
next section), by the constant factor $1+ \delta_{\tau}= \tau_n^{th}
/\tau_n^{ex}
= 1.008$, which should be regarded as an energy independent correction to
the weak process rates. This renormalization of the coupling guarantees the
correct prediction for $\tau_n$.

In Fig. 1 we report the Born rates $\omega_B$ for $n \leftrightarrow p$
processes, while in Fig. 2 we plot the corresponding radiative corrections,
$\Delta \omega_R/\omega_B$. Their effect is particularly large, up to $\sim
8 \%$, at low temperature.

\subsection{Finite nucleon mass corrections}

There are three additional contributions to the $n \leftrightarrow p$ rates
which appear when one relaxes the approximation of infinitely massive
nucleons. The
leading effects are proportional to $m_e/M_N$ or $T/M_N$, which, in the
temperature range relevant for BBN, can be as large as the radiative
corrections considered in the previous Section. This has been first pointed
out in \cite{Seckel} and then also numerically evaluated in \cite{EMMP1}.
At order $1/M_N$ there are new couplings appearing in the
expression of the weak hadronic current, the larger one coming from the
weak magnetic moments of nucleons
\be
J_\mu^{wm} = i \frac{G_F}{\sqrt{2}} \frac{f_2}{M_N} \, \ov{u}_p(p) \,
\sigma_{\mu \nu} \, (p-q')^\nu u_n(q')~~~,
\ee
where, from CVC, $f_2 = V_{ud} (\mu_p - \mu_n)/2= 1.81 V_{ud}$. Both scalar
and pseudoscalar contributions can be shown to be much smaller and
negligible for the accuracy we are interested in. At the same order in
inverse nucleon mass power it is also necessary to include the deformation
of the allowed phase space for the relevant scattering and decay processes,
due to nucleon recoiling. The sum of these two corrections 
for $n \leftrightarrow p$
rates with respect to th Born values, 
$\Delta \omega_M/\omega_B$, is plotted in Fig. 3.

The third effect is due to the initial nucleon thermal distribution. 
In the infinite nucleon mass limit, the average of weak rates over
nucleon distribution is in fact trivial, since the nucleon is at rest in
any frame. For finite $M_N$, by considering only $1/M_N$ terms, the effect
of the thermal average over the thermal spreading of the nucleon velocity 
produces a purely {\it kinetic} correction $\Delta
\omega_K$, whose expression can be reduced to a one dimensional integral
over electron momentum which can be numerically evaluated. The explicit
expression, which we do not report for brevity can be found in Section 4.2
and Appendix C of \cite{EMMP1}. The ratios $\Delta \omega_K /\omega_B$ for
$n \leftrightarrow p$ are reported in Fig. 4. Their size is rapidly growing
with temperature, since they are proportional to the ratio $T/M_N$.

\subsection{Thermal-Radiative corrections}

The $n \leftrightarrow p$ rates, calculated as the processes would occur in
vacuum, get slight corrections from the presence of the surrounding plasma
of $e^{\pm}$ pairs and $\gamma$. These are the so-called {\it
thermal-radiative effects}.

To compute these corrections one may use the standard real time formalism
for finite temperature field theory.  According to this scheme, field
propagators get additional contributions proportional to the number density
of that particular specie in the surrounding medium. For $\gamma$ and $e^{\pm}$
we have
\bea
i \, \Delta_\gamma^{\mu \nu}(k) &=& - \left[\frac{i}{k^2} + 2 \pi~
\frac{\delta(k^2)}{e^{\beta |k_0|} -1} \right] g^{\mu \nu} = -
\left[\frac{i}{k^2} + 2 \pi~ \delta(k^2)~B(k_0) \right] g^{\mu \nu}~~~,
\label{e:a1} \\
i \, S_e(p')&=& \frac{i}{\slp' - m_e} - 2 \pi~\delta({p'}^2 - m_e^2)~
F_e(p_0')~(\slp' + m_e)~~~.
\label{e:a2}
\eea
The entire set of thermal corrections $\Delta \omega_{TR}$, at first order
in its typical scale factor, i.e. $\alpha T/m_e$, have been computed by
several authors \cite{FTQFT} with quite different results. We have recently
performed this lengthy calculation in \cite{EMMP1}, to which we refer for
all details, and we have found a good agreement with the original result of
first reference in \cite{FTQFT}, namely that they contribute to correct the
Born rates only for less than $1 \%$.

\subsubsection{Radiative corrections on neutrino temperature}
\label{s:nut}

By assuming a sharp neutrino decoupling at $T_D = 2.3~MeV$ \cite{Enqvist},
the ratio $T_\nu / T$ can be evaluated using entropy conservation
\cite{EMMP1}. This leads to the expression
\be
\frac{T_\nu}{T} = \left\{ \begin{array}{cl} \left( \frac{{\cal I}(x_\gamma)
+ 2 {\cal I}(x_e) } { {\cal I}(x_\gamma^D) + 2 {\cal I}(x_e^D) }
\right)^{1/3} & T \leq T_D \\ 1 & T > T_D \end{array} \right.~~~,
\label{e:tnt}
\ee
with
\be
{\cal I}(x) =  \int_0^\infty \left(y^2 + 2 y x \right)^{1/2} \left( 4 y^2 +
8 y x  + 3 x^2 \right) \left[ \exp (x+y) {\pm} 1 \right]^{-1} ~dy~~~.
\label{e:int}
\ee
According to our notation $x_\alpha \equiv m_\alpha^R /T$ and $x_\alpha^D
\equiv m_\alpha^R/T_D$ with $\alpha=\gamma,e$ ($+$ or $-$ in the above
integrand is for fermions or bosons, respectively). Note that $m_\gamma^R$
and $m_e^R$ are the effective masses that photons and $e^{\pm}$ acquire in the
heat bath due to their interactions with the background plasma (see
Appendix A for details).

In \cite{EMMP1} the neutrino temperature $T_\nu$ as a function of photon
temperature $T$ was evaluated by using in (\ref{e:tnt}) the approximated
expressions $m_\gamma^R \simeq 0$ and $m_e^R \simeq m_e + \alpha T^2/ m_e$. This
simplified expression for $T_\nu$ has been used in all previous sections in
order to obtain the Born rates and their corrections as a function of $T$
only. The difference between the neutrino temperature
evaluated with the correct renormalized masses (\ref{e:mg}),
(\ref{e:me}), and the one obtained with the simplified
expressions results to be smaller than $0.01 \%$.
The corresponding effect on the rates (\ref{e:reaction}) due to this small
change in neutrino/photon temperature ratio, which can be
seen as a further sub-leading thermal-radiative correction to Born rates,
can be neglected at the level of precision we are interested in.

All thermal-radiative corrections to Born rates
$\Delta\omega_{TR} / \omega_B$ are reported in Fig. 5 . As evident
from this plot, around the
freeze--out temperature $T \sim 1 ~MeV$, $\Delta\omega_{TR}$ only
contribute as $\sim 0.4\%$ to the total rates. Thus they are clearly
subdominant. Note that changing $T_D$ in the range $(2 \div 3) MeV$ only
affects $T_\nu/T$ for less than $0.2 \%$.

\subsection{The total rates for $n \leftrightarrow p$ reactions}

In Fig. 6 we report the total corrections $\Delta \omega$ to Born rates 
$\omega_B$. In order to use
the corrected $n \leftrightarrow p$ rates $\omega = \omega_B + \Delta
\omega$
in the BBN code, it is useful to fit their
expressions as a function of the adimensional inverse photon temperature
$z\equiv m_e/T$,
\bea
\omega(n \rightarrow p) &=& \frac{1}{\tau_n^{ex}} \exp \left( -q_{np}
~z\right) ~\sum_{l=0}^{13} a_l ~z^{-l}~~~,~~~~~0.01~ MeV \leq T \leq 10 ~MeV
\label{e:fitnp} \\
\omega(p \rightarrow n) &=& \left\{ \begin{array} {cc}
\frac{1}{\tau_n^{ex}} \exp \left( -q_{pn}
~z\right) ~\sum_{l=1}^{13} b_l ~z^{-l} & 0.1~ MeV \leq T \leq 10 ~MeV \\
0 & 0.01~ MeV \leq T < 0.1 ~MeV \end{array} \right.
\label{e:fitpn}
\eea
with
\bea
\begin{array}{lll}
a_0= 1 & a_1= 0.160615 & a_2= 0.456817{\cdot} 10^1 \\
a_3= -0.401109 {\cdot} 10^2 & a_4= 0.137254 {\cdot} 10^3 & a_5=
-0.583644 {\cdot} 10^2 \\
a_6= 0.655938 {\cdot} 10^2 & a_7= -0.162185 {\cdot} 10^2 & a_8=
0.371109 {\cdot} 10^1 \\
a_9= -0.378497 & a_{10}= 0.223840 {\cdot} 10^{-1} & a_{11}= 0.723091 {\cdot}
10^{-5} \\
a_{12}= -0.462476 {\cdot} 10^{-4} & a_{13}= 0.186287 {\cdot} 10^{-5} &
q_{np}= 0.340994~~~, \\ \\
b_1= 0.199695 {\cdot} 10^2 & b_2= -0.671993 {\cdot} 10^2 & b_3=
0.109230 {\cdot} 10^3 \\
b_4= -0.295891 {\cdot} 10^1 & b_5= 0.407831 {\cdot} 10^2 & b_6=
-0.225830 {\cdot} 10^1 \\
b_7= 0.146751 & b_8= -0.185408 {\cdot} 10^{-2} & b_9= -0.205210 {\cdot}
10^{-3} \\
b_{10}= 0.158424 {\cdot} 10^{-5} & b_{11}= 0.369573 {\cdot} 10^{-6} & b_{12}=
-0.130731 {\cdot} 10^{-9} \\
b_{13}= -0.329060 {\cdot} 10^{-9} & q_{pn}= 2.89858 ~~~. &
\end{array}
\label{e:coeffnp} \\
\label{e:coeffpn}
\eea
The fit has been obtained requiring that the fitting functions differ by
less than $0.1 \%$ from the numerical values, while it is also
a good approximation to consider a vanishing rate $\omega(p \rightarrow n)$
for $T \leq 0.1 ~MeV$, see Eq. (\ref{e:fitpn}),
since it is a rapidly decreasing function with $T \rightarrow 0$.

\section{The set of equations for BBN }
\setcounter{equation}0

Let us consider $N_{nuc}$ species of nuclides, whose number densities,
$X_i=n_i/n_B$, are normalized with respect to the total baryon density
$n_B$.
The different nuclides are ordered in the following way: $n$, $H$,
$D$, $^3H$, $^3He$, $^4He$, $^6Li$, $^7Li$, $^7Be$, ... (for the complete
list see Ref. \cite{Kawano}).
Denoting with $R(t)$ the universe scale factor, the BBN set of
equations as functions of $R$, $n_B$, $T$, $X_i$, and of the electron
chemical potential $\phi_e \equiv \mu_e/T$ reads\footnote{We are using
natural units $\hbar=c=k_B=1$.}
\bea
\frac{\dot{R}}{R}  & =&  \sqrt{\frac{8 \pi}{3
M_P^2}}~\left[\rho_\gamma+\rho_e+\rho_\nu  +\rho_B\right]^{1/2}~~~,
\label{e:s1} \\
\frac{\dot{n}_B}{n_B} & = & - \, 3 \,\frac{\dot{R}}{R} = - \sqrt{\frac{24
\pi}{ M_P^2}}~\left[\rho_\gamma +\rho_e+\rho_\nu+\rho_B\right]^{1/2}~~~,
\label{e:s2}\\
L\left(\frac{m_e}{T},\phi_e\right) & = & \frac{\pi^2}{2} \frac{n_B}{T^3}
\sum_j Z_j X_j ~~~,
\label{e:s3} \\
\dot{T} & = & -\left\{3\frac{\dot{R}}{R} \left[\rho_\gamma \,    + \,
\p_\gamma \, + \, \rho_e \, + \, \p_e \, + \, \Theta(T-T_D) \, (\rho_\nu \,
+ \, \p_\nu)   \, + \, \p_B \right]    \; \right. \nonumber \\
& + & \left.  \frac{\partial \rho_e}{\partial \phi_e}\left(\sum_j
\frac{\partial \phi_e}{\partial X_j} \,\dot{X}_j - \, 3 \,
\frac{\dot{R}}{R}\,  n_B \,\frac{\partial \phi_e}{\partial n_B}\right)  \;
+ \; n_B \,\sum_j    \, \left(\Delta \mj \, + \, \frac{3}{2} T    \right)
\,\dot{X}_j \right\}  \nonumber \\
& {\times} & \left\{ \frac{d\rho_\gamma}{dT} \;    + \; \frac{\partial
\rho_e}{\partial T} \, + \,  \frac{\partial \rho_e}{\partial \phi_e} \,
\frac{\partial \phi_e}{\partial T}  \; + \; \Theta(T-T_D)
\,\frac{d\rho_\nu}{dT} \; + \; \frac{3}{2}    n_B \, \sum_j \, X_j
\right\}^{-1}
\label{e:s4}\\
\dot{X}_i & = & \sum_{j,k,l} \, N_i \left(  \Gamma_{kl \rt ij} \,
\frac{X_l^{N_l} \, X_k^{N_k}}{N_l! \, N_k !}  \; - \; \Gamma_{ij \rt kl} \,
\frac{X_i^{N_i} \, X_j^{N_j}}{N_i ! \, N_j  !} \right) \equiv
\Gamma_{i}(X_j)~~~.
\label{e:s5}
\eea
In the previous relations $\rho$ and $\p$ denote the energy density and the
pressure of an arbitrary particle specie. The function $L(z,y)$ in
(\ref{e:s3}) is defined as
\be
L(z,y)\equiv\frac 12 \int_z^\infty dx~x \,
\sqrt{x^2-z^2}\left(\frac{e^y}{e^x+e^y} - \frac{e^{-y}}{e^x+e^{-y}}\right)
~~~,
\label{e:s6}
\ee
$i,j,k,l=1,..,N_{nuc}$, and the $i$-th nuclide, with charge and atomic
number $(Z_i,A_i)$, has mass $M_i$ and mass excess ($M_u$ is the atomic
mass unit)
\be
\Delta M_i = M_i \, - \, A_i M_u~~~.
\ee
Moreover, in (\ref{e:s5}) we are considering in the sum a reaction between
$N_i$ nuclides of type $i$ and $N_j$ of type $j$ which results in $N_l$
nuclides of type $l$ and $N_k$ of type $k$, with its reverse reaction. The
energy density and the pressure of baryons take the form
\bea
\rho_B & = & n_B ~\left[M_u + \sum_j \left( \Delta \mj \, + \, \frac{3}{2}
T    \right) \, X_j \right]~~~,
\label{e:s6a}\\
\p_B & = & n_B ~T ~\sum_j X_j~~~.
\label{e:s6b}
\eea
Eq. (\ref{e:s1}) is easily recognized as the Friedmann equation where we
have neglected for simplicity the cosmological constant. Eq. (\ref{e:s2})
rules the scaling on $n_B$, whereas (\ref{e:s3}) states the neutrality of
primordial plasma. From entropy conservation one gets (\ref{e:s4}), and
(\ref{e:s5}) are the Boltzmann equations for the $N_{nuc}$ nuclide number
densities. Note that the presence of the $\Theta$-function in (\ref{e:s4})
is connected with neutrino decoupling at $T = T_D$.\footnote{We have
assumed that all neutrinos decouple at the same temperature $T_D$. Actually
muon and tau neutrinos decouple at a slightly larger temperature of $3.5 ~MeV$,
but nevertheless our approximation is largely consistent with the required
precision on $^4He$ yields.}

In the set of equations (\ref{e:s1})-(\ref{e:s5}) one can safely substitute
Eq. (\ref{e:s3}) with an analogous relation, obtained expanding the l.h.s.
of (\ref{e:s3}) with respect to $\phi_e$,
\be
L(z,y)\simeq y \, \int_z^\infty dx~x \, \sqrt{x^2-z^2} \,
\frac{e^x}{\left(e^x+1\right)^2} \equiv y \, f^{-1}(z)~~~.
\label{e:s7}
\ee
In this case, Eq. (\ref{e:s3}) provides an explicit expression for
$\phi_e=\phi_e(T,n_B,X_j)$,
\be
\phi_e \simeq \frac{\pi^2}{2}\, f\left( \frac{m_e}{T}\right)
\,\frac{n_B}{T^3} \sum_j Z_j X_j~~~.
\label{e:s8}
\ee
The consistency of this approach has been tested by means of an
iterative check.

The set of equations (\ref{e:s1})-(\ref{e:s5}) can be transformed in a set
of $N_{nuc}+1$ differential equations with the dimensionless variable $z=m_e/T$
as the evolution parameter. For numerical reasons, it is also better to
turn the variable $n_B$ into the dimensionless quantity $\hh \equiv
n_B/T^3$, which varies more slowly with $z$ than $n_B$. In terms of these
new variables the BBN set of equations becomes
\be
\frac{d \hh}{dz} \, = \, \left[ 1-\hH(z,\hh,X_j) ~G(z,\hh,X_j) \right] \frac{3\hh}{z}
~~~,
\label{e:basic1}
\ee
\be
\frac{dX_i}{dz} \, = \, G(z,\hh,X_j)~ \frac{\hgi}{z}~~~,
\label{e:basic2}
\ee
where the function $G$ is
\be
G(z,\hh,X_j) = \left\{ \frac{\sum_\alpha (4 \hrho_\alpha - z \frac{\partial
\hrho_\alpha}{\partial z}) + 4 \Theta(z_D-z) \hrho_\nu + \frac 32 \hh
\sum_j X_j}{3 \left[ \sum_\alpha (\hrho_\alpha + \hp_\alpha) + \frac 43
\Theta(z_D-z) \hrho_\nu + \hh \sum_j X_j \right] \hH + \hh \sum_j \left( z
\hdmj + \frac 32 \right) \hgj} \right\}.
\ee
In the previous equations $z_D=m_e/T_D$, $\alpha=e,\gamma$, and we have
considered the dimensionless Hubble parameter $\hH=H/m_e$,
\be
\hH(z,\hh,X_j) \, = \, \sqrt{\frac{8 \pi}{3}} \frac{m_e}{M_{P}} \frac{1}{z^2}
\left[\hrho_\gamma + \hrho_e + \hrho_\nu + \hh \left( z \hmu + \sum_j
\left( z \hdmj + \frac 32 \right) X_j \right) \right]^{1/2}~~~,
\label{e:hatHubble}
\ee
and the quantities $\hmu =M_u/m_e$, $\hdmj = \Delta M_j/m_e$, $\hgj =
\Gamma_j/m_e$. Energy densities and pressures
have been adimensionalized dividing by $T^4$. In Eq.s (\ref{e:basic1})
and (\ref{e:basic2}) we have neglected the terms containing the derivatives
of chemical potential.
In Appendix A we report the expressions for
$\hp_\alpha$ and $\hrho_\alpha$
evaluated taking into account the $\gamma$ and $e^{\pm}$
electromagnetic mass renormalization. As already mentioned in the previous
section this effect, changing the $\gamma$ and $e^{\pm}$ equations of
state, slightly modifies the $T_\nu/T$ ratio too. In 
order to speed up the numerical computation a fit of $\hp_\alpha$ and
$\hrho_\alpha$ as functions of $z$ has been performed and is also reported in
Appendix A.

The initial value for
(\ref{e:basic1}) is provided in terms of the final
baryon to photon density ratio $\eta$ according to the equation
\be
\hh_{in} = \frac{2 \zeta(3)}{\pi^2} \eta_{in} = \frac{11}{4}~ \frac{2
\zeta(3)}{\pi^2} \eta~~~.
\ee
The condition of Nuclear Statistical Equilibrium (NSE), which is
satisfied with high accuracy at the initial temperature $T_{in}=10 ~MeV$, is
then fixing the initial nuclide relative abundances. From NSE one gets,
for an arbitrary $i$-th nuclide with $g_i$ internal degrees of freedom,
\be
X_i(T_{in}) = \frac{g_i}{2} \left( \zeta(3)\sqrt{\frac{8}{ \pi}}
\right)^{A_i-1} A_i^{\frac{3}{2}} \, \left( \frac{T_{in}}{M_N}
\right)^{\frac{3}{2} (A_i-1)} \eta^{A_i-1} \, X_p^{Z_i} \, X_n^{A_i-Z_i} \,
\exp\left\{\frac{B_i}{T_{in}}\right\}~~~,
\label{e:s9}
\ee
where $B_i$ denotes the binding energy.

\section{Numerical Method}
\setcounter{equation}0

The most critical part of the BBN code concerns the method of numerical
resolution of the set of differential equations (\ref{e:basic1}),
(\ref{e:basic2}). In fact, since at high temperatures nuclear reactions
proceed in both forward and reverse directions with almost equal rapidity,
the r.h.s. of (\ref{e:basic2}) results to be a small difference of large
numbers. When this occurs the numerical problem is said {\it stiff}. As a
consequence, the step size is limited more severely by the requirement of
stability than by the accuracy of the numerical technique. In other words,
to preserve integration stability it is required to use a shorter step
size than what would be dictated by accuracy only. In order to manage the
problem, we use a NAG routine implementing a method belonging to the
class
of Backward Differentiation Formulas (BDFs) \cite{numrec}. 
This is quite a new approach
for BBN codes. In fact the standard code \cite{Wagoner,Kawano} uses instead the
implicit differentiating method (backward Euler scheme) \cite{numrec} for
writing the r.h.s. of (\ref{e:basic2}) and a Runge-Kutta solver.

\begin{table}[t]
\begin{center}
Table \ref{t:reac} \\
\vspace{.6truecm}
\begin{tabular}{|r|rcl|}\hline\hline
1) & $n$ & $\leftrightarrow$ & $p$ \\
2) & $T$ & $\leftrightarrow$ &$^3He$ \\
3) & $p~+~n$ & $\leftrightarrow$ & $D~+~\gamma$ \\
4) & $n~+~D$ & $\leftrightarrow$ & $T~+~\gamma$ \\
5) & $n~+~^3He$ & $\leftrightarrow$ & $^4He~+~\gamma$ \\
6) & $n~+~^6Li$ & $\leftrightarrow$ & $^7Li~+~\gamma$ \\
7) & $n~+~^3He$ & $\leftrightarrow$ & $T~+~p$ \\
8) & $n~+~^7Be$ & $\leftrightarrow$ & $^7Li~+~p$ \\
9) & $n~+~^6Li$ & $\leftrightarrow$ & $^4He~+~T$ \\
10) & $n~+~^7Be$ & $\leftrightarrow$ & $^4He~+~^4He$ \\
11) & $p~+~D$ & $\leftrightarrow$ & $^3He~+~\gamma$ \\
12) & $p~+~T$ & $\leftrightarrow$ & $^4He~+~\gamma$ \\
13) & $p~+~^6Li$ & $\leftrightarrow$ & $^7Be~+~\gamma$ \\
14) & $p~+~^6Li$ & $\leftrightarrow$ & $^4He~+~^3He$ \\
15) & $p~+~^7Li$ & $\leftrightarrow$ & $^4He~+~^4He$ \\
16) & $D~+~^4He$ & $\leftrightarrow$ & $^6Li~+~\gamma$ \\
17) & $T~+~^4He$ & $\leftrightarrow$ & $^7Li~+~\gamma$ \\
18) & $^3He~+~^4He$ & $\leftrightarrow$ & $^7Be~+~\gamma$ \\
19) & $D~+~D$ & $\leftrightarrow$ & $^3He~+~n$ \\
20) & $D~+~D$ & $\leftrightarrow$ & $T~+~p$ \\
21) & $D~+~T$ & $\leftrightarrow$ & $^4He~+~n$ \\
22) & $D~+~^3He$ & $\leftrightarrow$ & $^4He~+~p$ \\
23) & $^3He~+~^3He$ & $\leftrightarrow$ & $^4He~+~p~+~p$ \\
24) & $D~+~^7Li$ & $\leftrightarrow$ & $^4He~+~^4He~+~n$ \\
25) & $D~+~^7Be$ & $\leftrightarrow$ & $^4He~+~^4He~+~p$\\
\hline\hline
\end{tabular}
\end{center}
\caption{The reduced network of nuclear reactions.}
\label{t:reac}
\end{table}
Few comments on the different numerical methods are in order. Let us
consider the differential equation
\be
\frac{dy(t)}{dt} = f(t,y(t))~~~.
\ee
In the Runge-Kutta methods the solution at $t_{i+1}$ is completely
determined by its value at $t_i$ (one-step methods), namely the solver has
{\it no memory}. A different approach is provided by a wide class of
numerical methods
referred to as {\it multistep methods} like BDFs. Here, the values of the
solution at $t_k$ ($k=i,i-1,...,i-p$), $y(t_k)\equiv y_k$, previously
computed, and the unknown value $y(t_{i+1})\equiv y_{i+1}$, are
interpolated by a polynomial, $P(t;~y_{i+1},~y_i,...)$, in order to
approximate the solution and its derivative. Substituting in the
differential equation,
\be
\frac{dP}{dt} (t_{i+1};~y_{i+1},~y_i,...) \simeq f(t_{i+1},~y_{i+1})~~~,
\ee
one obtains a family of BDFs,
\begin{eqnarray}
(t_{i+1}-t_{i}) \, f(t_{i+1},~y_{i+1}) & \simeq &
P(t_{i+1};~y_{i+1},~y_i,...) \, - \,
P(t_i;~y_{i+1},~y_i,...) \nonumber \\
& = & \alpha_0 y_{i+1} + \alpha_1 y_i + ... ~~~.
\label{e:bdfs}
\end{eqnarray}
Two methods can be used for solving the previous equation in the implicit
case, $\alpha_0 \neq 0$: functional iteration and Newton's method. In the
former case some initial guess is taken for $y_{i+1}$ and refined by
iteration. In the latter case, one linearizes Eq. (\ref{e:bdfs}) by
expanding $f$ around $y_i$. The new point, $y_{i+1}$, is then found by
inverting a matrix, in a way similar to the backward Euler scheme.
The NAG routine implements both methods and incorporates an error control 
test, which drives the step-size adjustment.

The nuclear reaction network used in the code includes all the 88 reactions
between the 26 nuclides present in the standard code \cite{Wagoner,
Kawano}. We used the same nuclear rate data of the
standard code, which are collected and updated in \cite{network}. In order
to reduce the
computation time one can also choose a reduced network, made of the 25 reactions
between 9 nuclides listed in Table \ref{t:reac}. Using the complete network
we have verified that the reduced one affects the
abundances for no more than $0.01 ~\%$, while it greatly reduces the evaluation
time.

\section{Results on Light Element Abundances}
\setcounter{equation}0
\label{s:numres}

The reliable numerical code just discussed can now be used to study the
effect of the different corrections to $n \leftrightarrow p$ Born rates on
light elements abundances. By definition
\be
Y_2 = \frac{X_3}{X_2}~~~,~~~~~
Y_3 = \frac{X_5}{X_2}~~~,~~~~~
Y_4 = \frac{M_6~ X_6}{\sum_j M_j~ X_j} ~~~~, ~~~~~
Y_7 = \frac{X_8}{X_2} ~~~.
\ee
In the first two rows of Table \ref{t:abund} are shown 
the predictions for $Y_2$, $Y_3$, $Y_4$ and $Y_7$ at $\eta = 5 \cdot 10^{-10}$,
corresponding to the
complete $n \leftrightarrow p$ rates, $\omega_{Tot}$, and to the Born
approximation, $\omega_B$ \footnote{Note that, according to our notation,
with $\omega_B$ we denote the pure Born predictions for $n \leftrightarrow
p$ rates without any constant rescaling of coupling
to account for the experimental
value of neutron lifetime (see Section \ref{s:rad}).}.
As it is clear from Table \ref{t:abund}, the main effect of the
corrections, which results into the enhancement of the $n \rightarrow p$ conversion
rate, is to allow a smaller number of neutrons to survive till the onset of
nucleosynthesis. This ends up in a smaller fraction of elements which fix
neutrons with respect to hydrogen.

The effects on light element yields due to the various corrections with
respect to the Born predictions are also reported in Table \ref{t:abund}.
For all nuclides the pure radiative correction $\Delta \omega_R$ provides
the dominant contribution, while the finite nucleon mass effects, the
kinetic and the thermal-radiative ones almost cancel each other. Finally the
last row reports the further contribution due to the
additional term required to recover the experimental
neutron lifetime \cite{EMMP1}.

If we make use of the results of 
\cite{Fiorentini} to quantify the uncertainties coming from nuclear reaction
processes, we observe that only for $Y_4$ the radiative
correction affects the Born result by an amount larger than the theoretical
uncertainties, including nuclear reactions. For $^4He$ mass fraction in
fact, the theoretical uncertainty due to nuclear reaction rates is
estimated to be of the order of $0.1\%$ and thus comparable with the
uncertainty due to the experimental error on neutron lifetime. For $D$,
$^3He$ and $^7Li$ the uncertainty due to the poor knowledge of nuclear
reaction rates is estimated to be of the order of $(10 \div 30)\%$
\cite{Fiorentini}, thus completely covering any radiative/thermal
correction on $n \leftrightarrow p$ rates.

\begin{table}[t]
\begin{center}
Table \ref{t:abund} \\
\vspace{.6truecm}
\begin{tabular}{|c|c|c|c|c|} \hline\hline
& $Y_2$ & $Y_3$ & $Y_4$ & $Y_7$ \\
\hline\hline &&&& \\
$\omega_{Tot}$  & $0.3638{\cdot} 10^{-4}$ & $0.1175{\cdot} 10^{-4}$ &
$0.2446$ & $0.2814{\cdot} 10^{-9}$ \\
&&&&\\
$\omega_{B}$  & $0.3727{\cdot} 10^{-4}$ & $0.1184{\cdot} 10^{-4}$ &
$0.2550$ & $0.2873{\cdot} 10^{-9}$ \\
&&&&\\
\hline \hline &&&&\\
$\Delta \omega_R$  & $-2.3\%$ & $-2.8\%$ & $-3.8\%$ & $-1.9 \%$ \\
&&&&\\
$\Delta \omega_M$ & $.2\%$ & $.1\%$ & $.3\%$  & $.2 \%$ \\
&&&&\\
$\Delta \omega_K$ & $.2\%$ & $.1\%$ & $.3\%$ & $.2 \%$ \\
&&&&\\
$\Delta \omega_{TR}$  & $-.6\%$ & $-.1\%$ & $-.7\%$ & $-.4 \%$ \\
&&&&\\
$\delta_\tau \omega_T$  & $-.3\%$ & $-.1\%$ & $-.6\%$ & $-.3 \%$ \\
&&&&\\
\hline
\end{tabular}
\end{center}
\caption{The predictions on light element abundances obtained with the
numerical code for $\eta= 5 \cdot 10^{-10}$ and $N_\nu=3$.
In the lower rows the effect of the various corrections is
reported.}
\label{t:abund}
\end{table}
In Fig. 7 the predictions on $Y_4$ are shown versus $\eta$ for $N_\nu=2,3,4$ 
and for a
$1~\sigma$ variation of $\tau_n^{ex}$. The two experimental estimates for
the primordial $^4He$ mass fraction, $Y_4^{(l)}$ and $Y_4^{(h)}$, as
horizontal bands, are also reported. Fig.s 8 and 9 show the predictions for
$D$ and $^7Li$ abundances. Note that, due to the negligible variation of
$Y_2$ and $Y_7$ on small $\tau_n$ changes, no splitting of predictions for
$1~\sigma$ variation of $\tau_n^{ex}$ is present.

A fit, up to $1 \%$ accuracy, of the relevant observables $Y_2$,
$Y_3$, $Y_4$ and $Y_7$ as a function of $x =
\log_{10}\left(10^{10}\eta\right)$, $N_\nu$ and $\tau_n$ has been
performed. The following expressions have been obtained
\bea
10^{3} {\cdot} Y_2 & = & \left[ \sum_{i=0}^{4} a_i \, x^{i} + a_5 \,( N_\nu
-3)\right] \,\exp\left\{ a_6 \,x + a_7 \, x^2 \right\}~~~,
\label{e:abuy2}\\
10^{5}  {\cdot} Y_3 & = & \left[ \sum_{i=0}^{4} a_i \, x^{i} + a_5 \,(
N_\nu -3)\right] \,\exp\left\{ a_6 \,x \right\}~~~,
\label{e:abuy3} \\
10 {\cdot} Y_4 & = & \sum_{i=0}^{5} a_i \, x^{i} + a_6 \,(\tau - \tau_{ex})
+ a_7 \,( N_\nu -3)+ a_8 \, x \, (\tau - \tau_{ex}) +a_9 \, x\, (
N_\nu-3)~~~,
\label{e:abuy4} \\
10^{9} {\cdot} Y_7 & = & \left[ \sum_{i=0}^{3} a_i \, x^{i} + a_4 \,( N_\nu
-3)+ a_5 \, x \,( N_\nu -3)\right] \exp\left\{ \sum_{i=1}^4 a_{5+i} ~x^i
\right\}~~~, 
\label{e:abuy7}
\eea
where $\tau_{ex}= 886.7\, s$ and the values of the fit coefficients are
reported in Table \ref{t:coefit}.

Neutrino decoupling has been shown by many authors \cite{neutrdec} to be
a process which still takes place when $e^{\pm}$ pairs annihilate. This implies
that neutrinos are in fact slightly reheated during this annihilation process
and their final distribution in momentum space shows an interesting non
equilibrium shape. In Ref. \cite{Fields} it is estimated that the effect 
on $Y_4$
due to the inclusion of this slight neutrino heating is very small, $\delta Y_4
\sim 1.5~ 10^{-4}$, in the whole range $10^{-10} \leq \eta \leq 10^{-9}$.
We have included this constant correction to $Y_4$ prediction.

\begin{table}[t]
\begin{center}
Table \ref{t:coefit} \\
\vspace{.6truecm}
\begin{tabular}{|c|c|c|c|c|} \hline\hline
Coeff. & $ 10^{3} {\cdot} Y_2 $ & $10^{5} {\cdot} Y_3 $ & $10 {\cdot} Y_4 $
& $10^{9} {\cdot} Y_7$ \\
\hline\hline&&&&\\
$a_0$  & $ 0.4854$ & $ 3.325$ & $ 2.209$ & $ 0.5419$ \\ &&&&\\
$a_1$ & $ 0.2919$ & $ 0.1496$ & $ 0.5548$ & $ -0.5981$ \\ &&&&\\
$a_2$ & $-0.3516$ & $ 1.597$ & $-0.6491$ & $ -1.914$ \\ &&&&\\
$a_3$ & $ 0.5048$ & $-1.923$ & $ 0.7661$ & $ 4.521$ \\ &&&&\\
$a_4$  & $ -0.4269$ & $ 1.312$ & $ -0.5366$ & $ 0.1587$ \\ &&&&\\
$a_5$  & $  0.7772{\cdot} 10^{-1} $ & $ 0.1782$ & $ 0.1614$ & $-0.3256$\\
&&&&\\
$a_6$ & $ -4.397$ & $-1.705$ & $0.2059{\cdot} 10^{-2}$ & $ -4.102$ \\ &&&&\\
$a_7$ & $ 0.5925$ & $/$ & $ 0.1300$ & $ 5.072$ \\ &&&&\\
$a_8$  & $/$ & $/$ & $-0.4156{\cdot} 10^{-4}$ & $-1.209$ \\ &&&&\\
$a_9$  & $/$ & $/$ & $0.7433{\cdot} 10^{-2}$ & $-0.6269$ \\&&&&\\
\hline
\end{tabular}
\end{center}
\caption{Values of the fit coefficients (\ref{e:abuy2})-(\ref{e:abuy7})
for light element abundances.}
\label{t:coefit}
\end{table}
From the fit reported in Eq. (\ref{e:abuy4}) it is easy to quantify the 
theoretical error on $Y_4$. Since this is basically due to the uncertainty
on $\tau_n$ we have
\be
\frac{\Delta Y_4}{Y_4} = \frac{(a_6 + a_8 x) \Delta \tau_n}{10 Y_4}
\leq 0.1 \%~~~.
\label{uncy}
\ee
In Fig. 10, our prediction for $Y_4$ of Eq. (\ref{e:abuy4}) with $N_\nu=3$ and
$\tau_n=885.3~s$, is compared with an analogous fit, $Y_4'$, performed in
\cite{Sarkar}. The agreement between the two expressions
obtained by independent codes is up to $1 \%$ in the relevant range for
$\eta$. 

In Fig. 11 we present the results of a likelihood analysis of the
theoretical predictions obtained with the numerical code for the four
combinations of experimental results: a) high $D$, low $^4He$; b) high $D$,
high $^4He$; c) low $D$, low $^4He$; d) low $D$, high $^4He$, and using
the low value for $^7Li$ abundance.
In particular, we plot the product of gaussian distribution for $D$, $^4He$ 
and $^7Li$ centered around the measured values and with their 
corresponding experimental errors,
\bea
L(N_\nu,x) &=& \exp\left\{-{(Y_2(N_\nu,x)- Y_2^{ex})^2}\over{2 \sigma^2_{2}}
\right\} \exp\left\{-{(Y_4(N_\nu,x)- Y_4^{ex})^2}\over{2 \sigma^2_{4}} \right\} 
\nonumber \\
&\times&\exp\left\{-{(Y_7(N_\nu,x)- Y_7^{ex})^2}\over{2 \sigma^2_{7}} 
\right\}~~~. 
\label{e:likely}
\eea
Notice that all functions have been normalized to unity in the maximum.
As is clear from the plots, the analysis prefers the high value of
$D$ (plots a) and b)). In both cases the distributions are centered in the
range $x\in 0.2\div 0.4$, but at $N_\nu\sim 3$ for low $^4He$ and
$N_\nu\sim 3.5$ for high $^4He$. For low $D$ the compatibility with
experimental data is worse.  Note that c) and d) distributions have been
multiplied by a factor of 25 and 100 respectively and centered in the
range $x\in 0.6\div 0.8$, and at $N_\nu\sim 2$ for low $^4He$ and
$N_\nu\sim 3$ for high $^4He$. The better agreement at $1 \sigma$ of the
data set a) and b) with the theoretical predictions is basically due to
the effect of $^7Li$ data which corresponds to values for $\eta$ compatible
with low $D$ data of c) and d) at $2 \sigma$ only.
It should be mentioned however that these 
results only take into account experimental errors, so that the confidence 
level regions in the $N_\nu - x$ plane would be broader by convoluting the 
considered distributions with the ones containing the theoretical error. 

\section{Conclusions}
\setcounter{equation}0

In this paper a detailed study of the effects on light element yields of
the radiative, finite nucleon mass, thermal and plasma corrections to Born
rates (\ref{e:reaction}) has been carried out. The aim of such an analysis
was to reduce the error on, basically, $Y_4$ to less than $1 \%$, which is
motivated by the most recent experimental determinations for $^4He$ abundance.
This accurate analysis has
been performed using an update version of the BBN standard code
\cite{Wagoner,Kawano}. A different numerical approach, based on BDF techniques
has been implemented to solve the stiff Boltzmann equations for nuclei
densities.
The numerical results for $^4He$ mass fraction almost
confirm the computation reported in Ref. \cite{Sarkar}, while the theoretical
error, also including the propagation of uncertainties on nuclear processes, 
as estimated in \cite{Fiorentini}, is of the order of $0.1 \%$.
Our
analysis shows that the preferred experimental values are high value for
$D$ and low one for $^4He$, in which case the distribution is centered at 
$x\sim 0.3$ and $N_\nu \sim 3$.

\newpage

\appendix

\section{ Radiative corrections to $e^{\pm}$, $\gamma$ equations of state}
\setcounter{equation}0

In an accurate description of the primordial plasma it is important
to consider the electromagnetic correction to the
$e^{\pm}$ and $\gamma$ equations of state induced by the $e^{\pm}$ and
$\gamma$ mass renormalization.

As well known, the photon renormalized mass, up to first order
correction in the electromagnetic coupling constant $\alpha$, reads
\cite{photon}
\be
m_\gamma^R(z) \simeq m_e \, \frac 2z \,\sqrt{\frac{\alpha }{\pi}} \,
\left[\int_z^\infty \, d x~\frac{\sqrt{x^2 \, - \,  z^2}}{1+e^x
}\right]^{1/2}~~~,
\label{e:mg}
\ee
and for $e^{\pm}$ \cite{FTQFT}
\bea
m_e^R(z,y) \simeq m_e\left\{1+\frac{\alpha}{\pi z^2} \left[ \frac{\pi^2}{3}
+ \int_z^\infty \left( 2 \sqrt{x^2-z^2} + \frac{z^2}{2 \sqrt{y^2-z^2}}
\log\Lambda \right)\, \frac{dx}{1+e^x} \right]\right\}~~~,
\label{e:me}
\eea
where $z\equiv m_e/T$, $y\equiv E_e/T$ and
\be
\Lambda(x,y,z)=\frac{x^2y^2-\left(z^2+ \sqrt{x^2-z^2}
\sqrt{y^2-z^2}\right)^2} {x^2 y^2 -\left(z^2- \sqrt{x^2-z^2}
\sqrt{y^2-z^2}\right)^2}~~~.
\label{e:lambda}
\ee
Note that in Eq.s (\ref{e:mg}) and (\ref{e:me}) one can neglect the
contribution of electron chemical potential, $\phi_e \equiv \mu_e/T$, due
to its small value.

By using (\ref{e:mg}) and (\ref{e:me}) in the expressions of $\rho_\gamma$,
$\p_\gamma$, $\rho_e$, $\p_e$ one gets the latter quantities as
functions of $z$ only. Since the $e^{\pm}$ and $\gamma$ energy densities
and pressures have to be used in a BBN code, in order to speed up the
computation one can fit these quantities as function of $z$ and use these
fits in the evolution equations. The fitted expressions for the
dimensionless electron energy density and pressure, $\hrho_e=\rho_e/T^4$
and $\hp_e=\p/T^4$, in the range $z\in [0.05,8.52]$ ($\hrho_e=\hp_e=0$
for $z>8.52$), result to be
\bea
\hrho_e(z) & = & 1.145 + 0.33981{\cdot} 10^{-1} z - 0.14543 z^2 +
0.25507{\cdot} 10^{-1} z^3 - 0.54168{\cdot} 10^{-3} z^4 \nonumber \\
& - & 0.11263{\cdot} 10^{-3} z^5 - 0.29742{\cdot} 10^{-5} z^6 +
0.38331{\cdot} 10^{-6} z^7 + 0.45263{\cdot} 10^{-7} z^8 \nonumber \\
&+& 0.19241{\cdot} 10^{-8} z^9 - 0.96597{\cdot} 10^{-10} z^{10} -
0.19505{\cdot} 10^{-10} z^{11} -0.14079{\cdot} 10^{-12} z^{12}, \nonumber\\
\label{e:fit1} \\
\hp_e(z) & = & \left( 0.3786 +0.19126{\cdot} 10^{-1}z -0.63895{\cdot}
10^{-1}z^2 +0.32085{\cdot} 10^{-1}z^3 -0.48501{\cdot} 10^{-2}z^4\right.
\nonumber \\
&-& 0.16611{\cdot} 10^{-3}z^5 + 0.82922{\cdot} 10^{-4} z^6 + 0.79884{\cdot}
10^{-5} z^7 - 0.60619{\cdot} 10^{-6} z^8 \nonumber \\
&-& \left. 0.19568{\cdot} 10^{-6} z^9 -0.10921{\cdot} 10^{-7} z^{10} +
0.38564{\cdot} 10^{-8} z^{11} \right) ~ e^{-0.13145~z^2} ~~~. \nonumber \\
\label{e:fit2}
\eea
Moreover, in the considered temperature range, one can show that
$\hrho_\gamma=\rho_\gamma/T^4$ only varies between 0.6580 and 0.6573, and
$\hp_\gamma=\p_\gamma/T^4$ between 0.2193 and 0.2187. Thus, for simplicity,
one can assume $\hrho_\gamma$ constant and equal to the average value
0.6577, and to 0.2190 for $\hp_\gamma$.

\begin{figure}
\epsffile{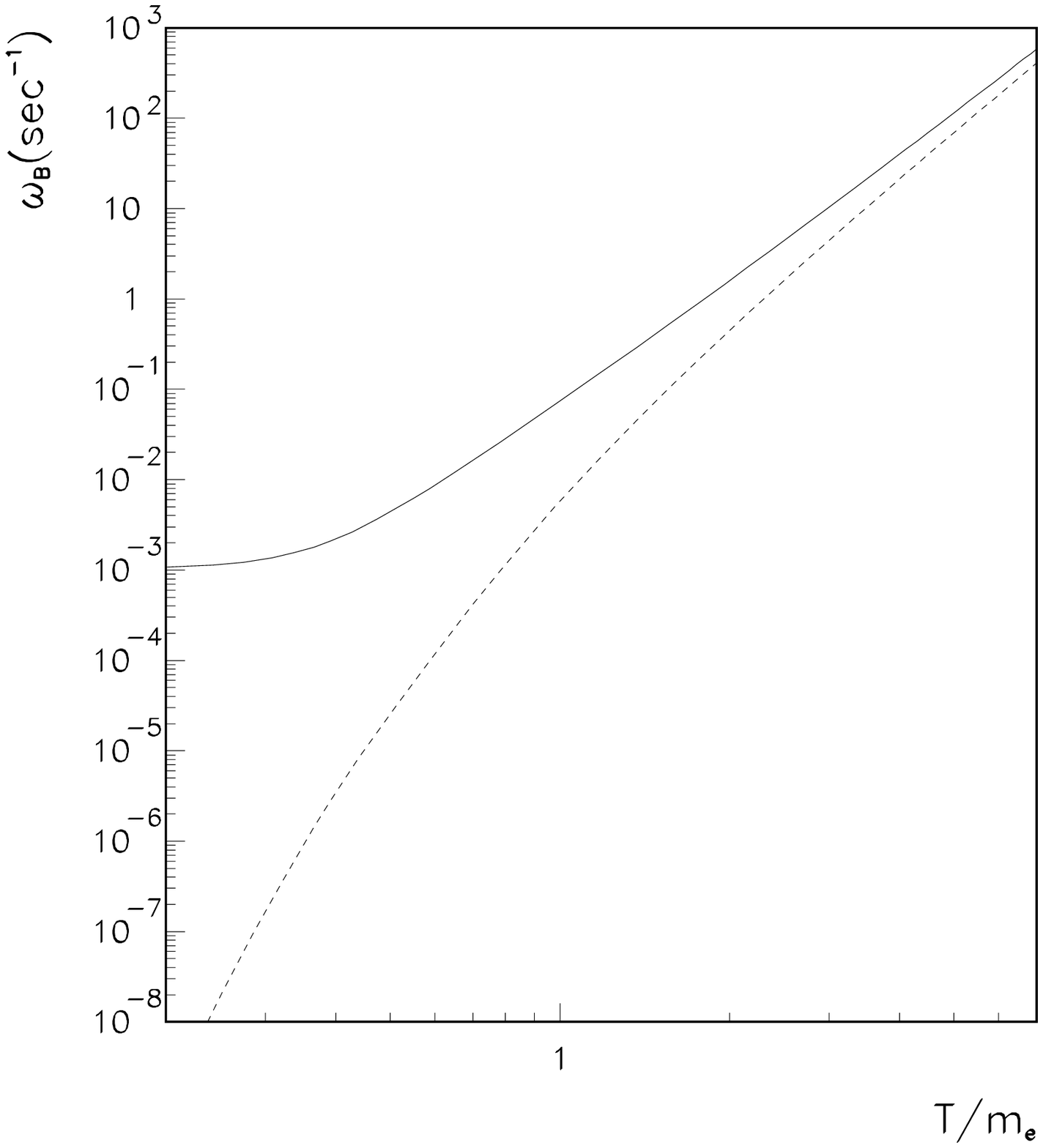}
\caption{The total Born rates, $\omega_B$, for $n \rightarrow p$ (solid
line) and $p \rightarrow n$ transitions (dashed line). This
notation is adopted hereafter.}
\end{figure}
\begin{figure}
\epsffile{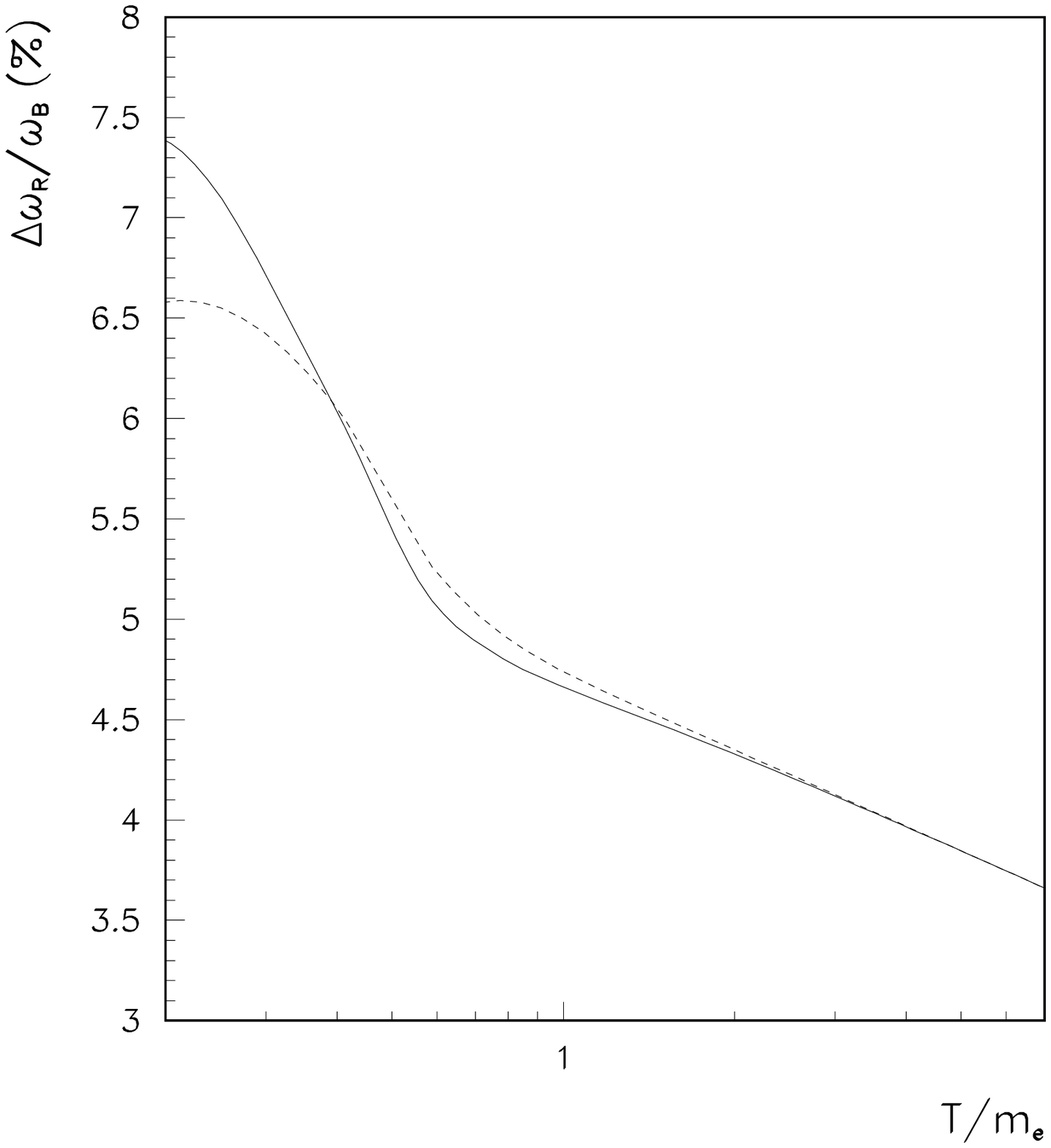}
\caption{The radiative corrections to Born rates, $\Delta
\omega_R/\omega_B$, for $n \leftrightarrow p$ transitions.}
\end{figure}
\begin{figure}
\epsffile{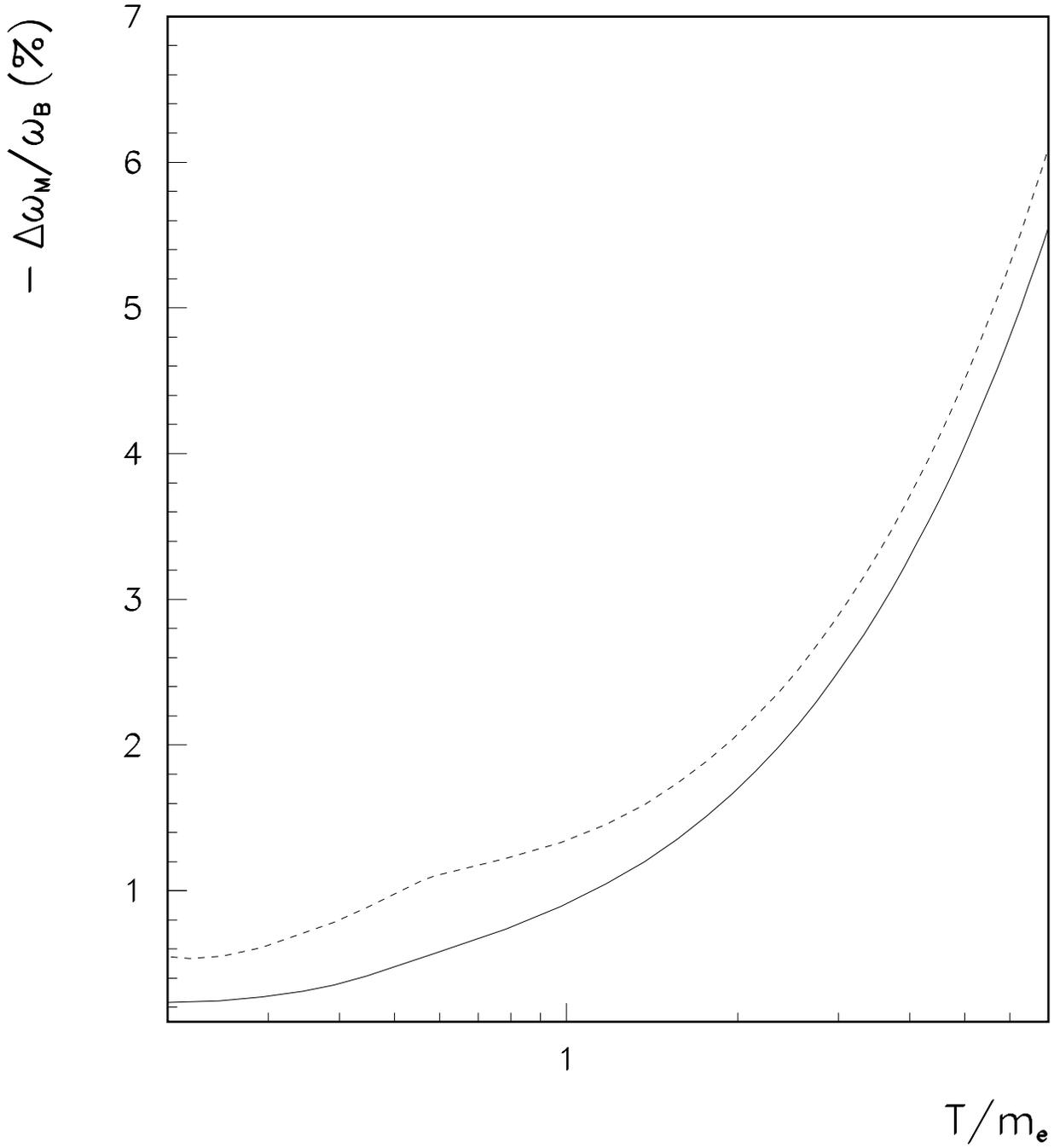}
\caption{The finite nucleon mass corrections to Born rates, $\Delta
\omega_M/\omega_B$, for $n \leftrightarrow p$ transitions.}
\end{figure}
\begin{figure}
\epsffile{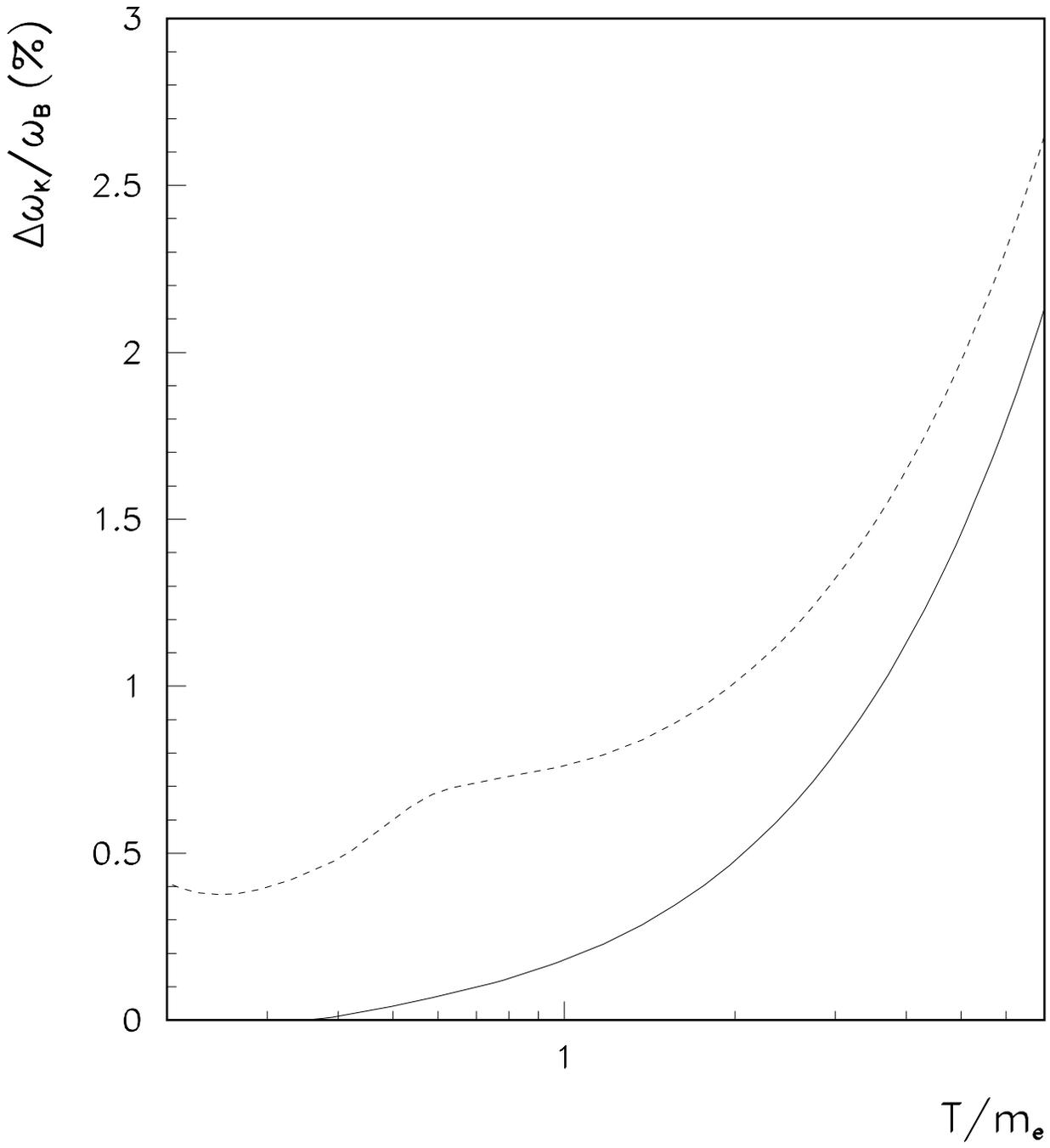}
\caption{The {\it kinetic} corrections to Born rates, $\Delta
\omega_K/\omega_B$, for $n \leftrightarrow p$ transitions.}
\end{figure}
\begin{figure}
\epsffile{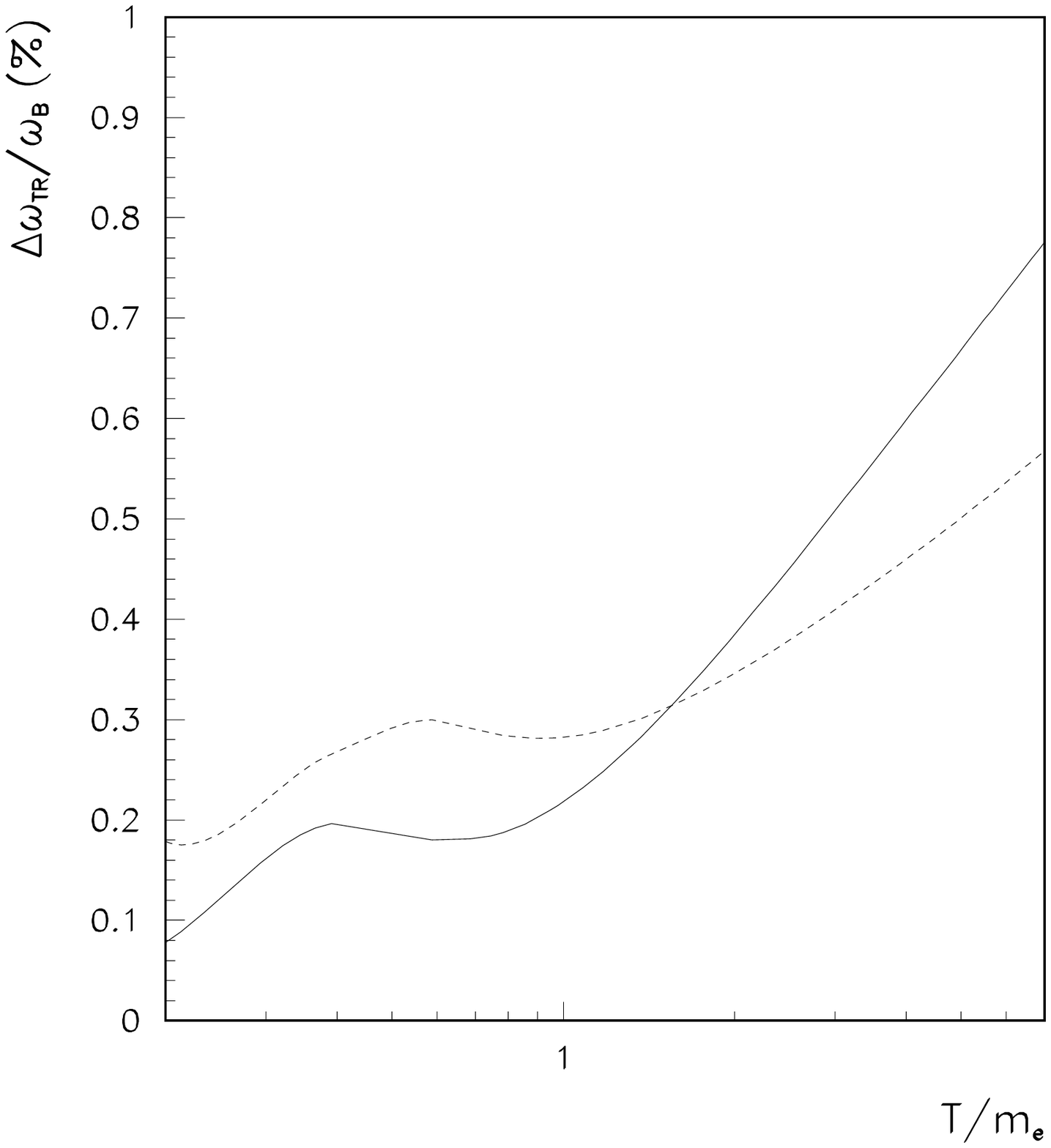}
\caption{The thermal-radiative corrections to Born rates, $\Delta
\omega_{TR} /\omega_B$, for $n \leftrightarrow p$ transitions.}
\end{figure}
\begin{figure}
\epsffile{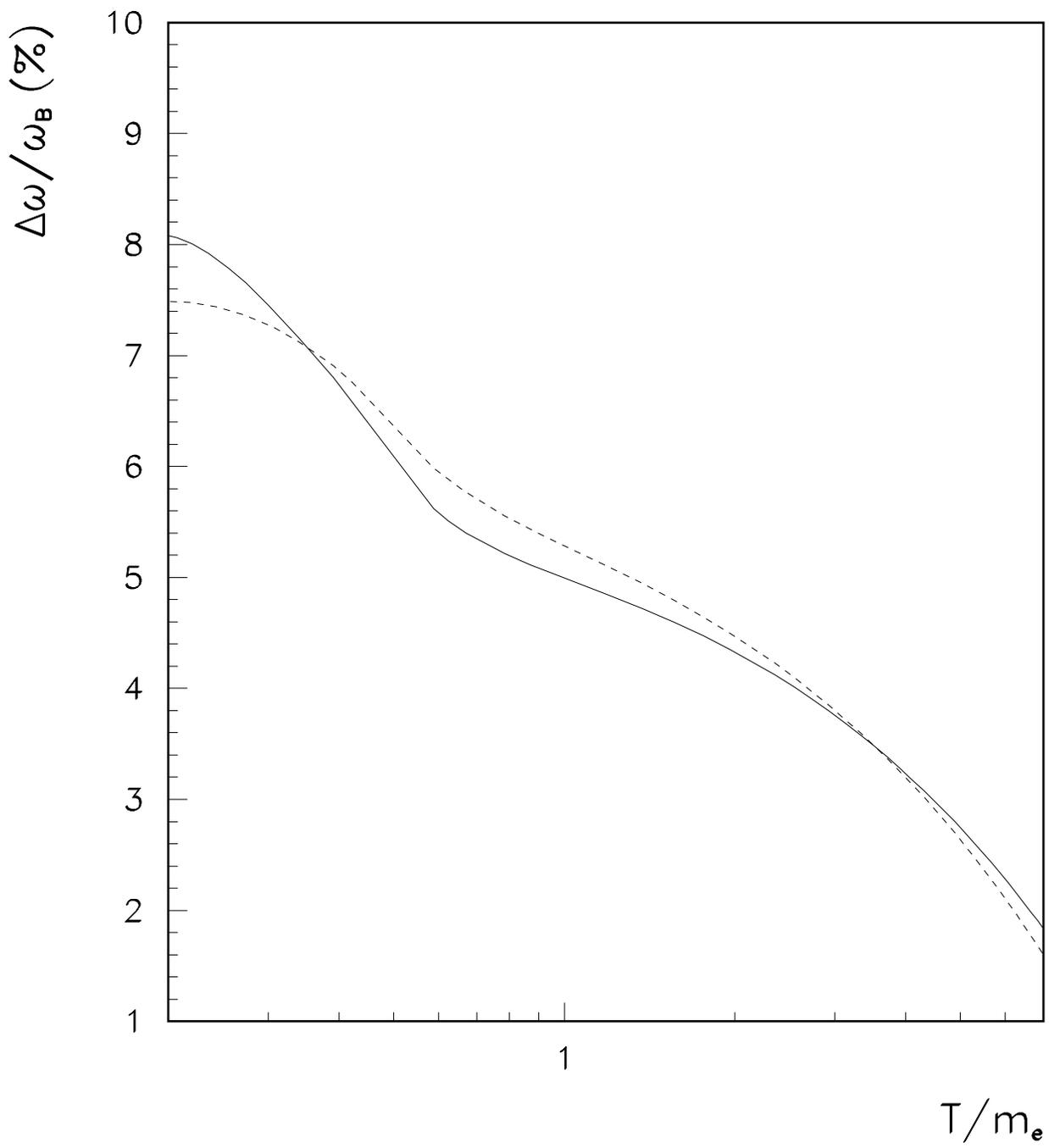}
\caption{The total corrections to Born rates for $n \leftrightarrow p$
transitions.}
\end{figure}
\begin{figure}
\epsffile{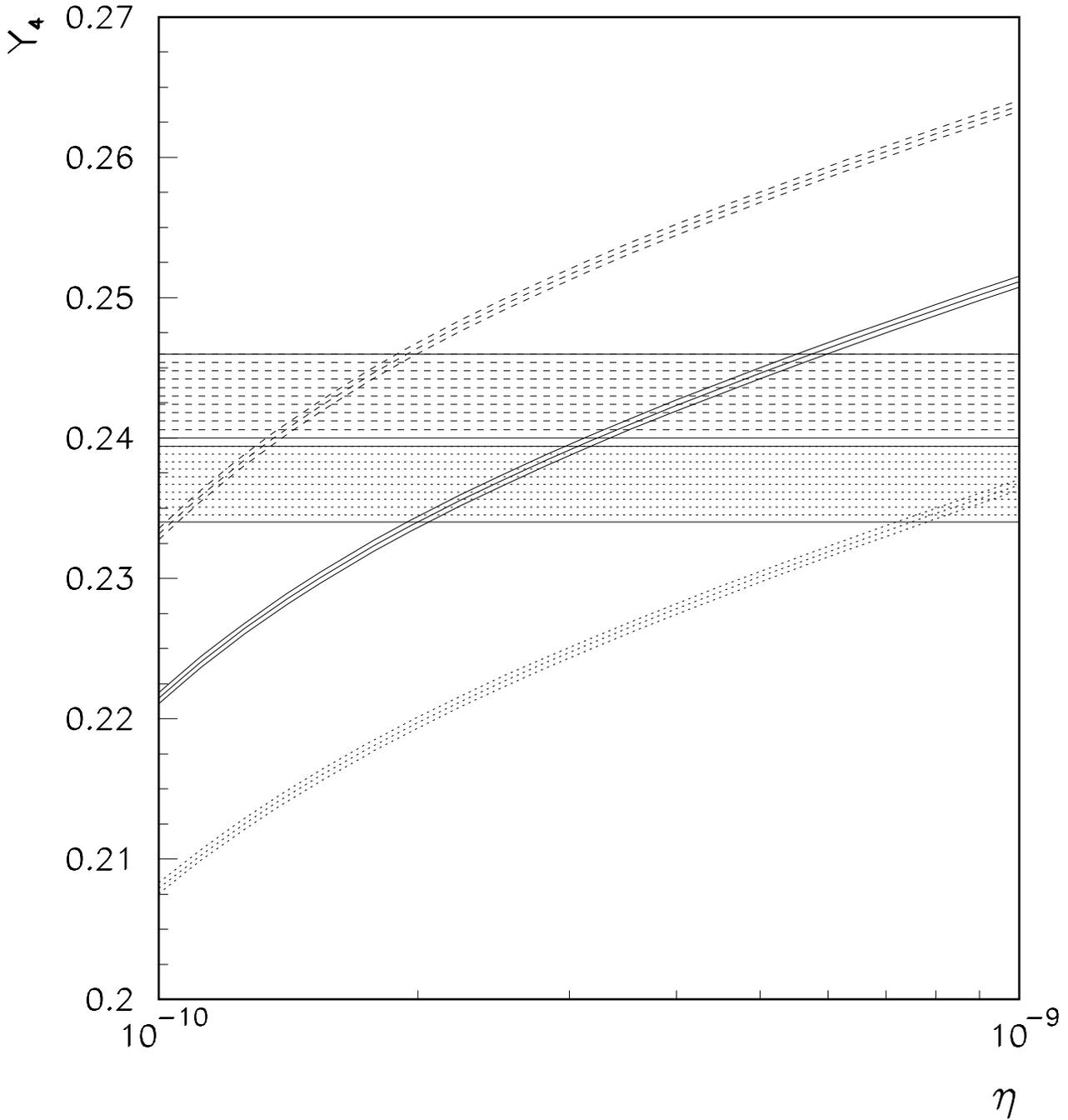} \vspace{-2cm}
\caption{The $^4He$ mass fraction, $Y_4$, versus $\eta$. The three
solid lines are, from larger to lower values of $Y_4$, the predictions
corresponding to $N_\nu=3$ and $\tau_n^{ex}=888.6~s$, $886.7~s$, $884.8~s$,
respectively. Analogously, the dashed lines correspond to $N_\nu=4$ and the
dotted ones to $N_\nu=2$. The dotted and dashed horizontal band are the
experimental values of Ref.s \protect\cite{Steigman} and \protect\cite{Izotov},
respectively, with $1 \sigma$ interval.}
\end{figure}
\begin{figure}
\epsffile{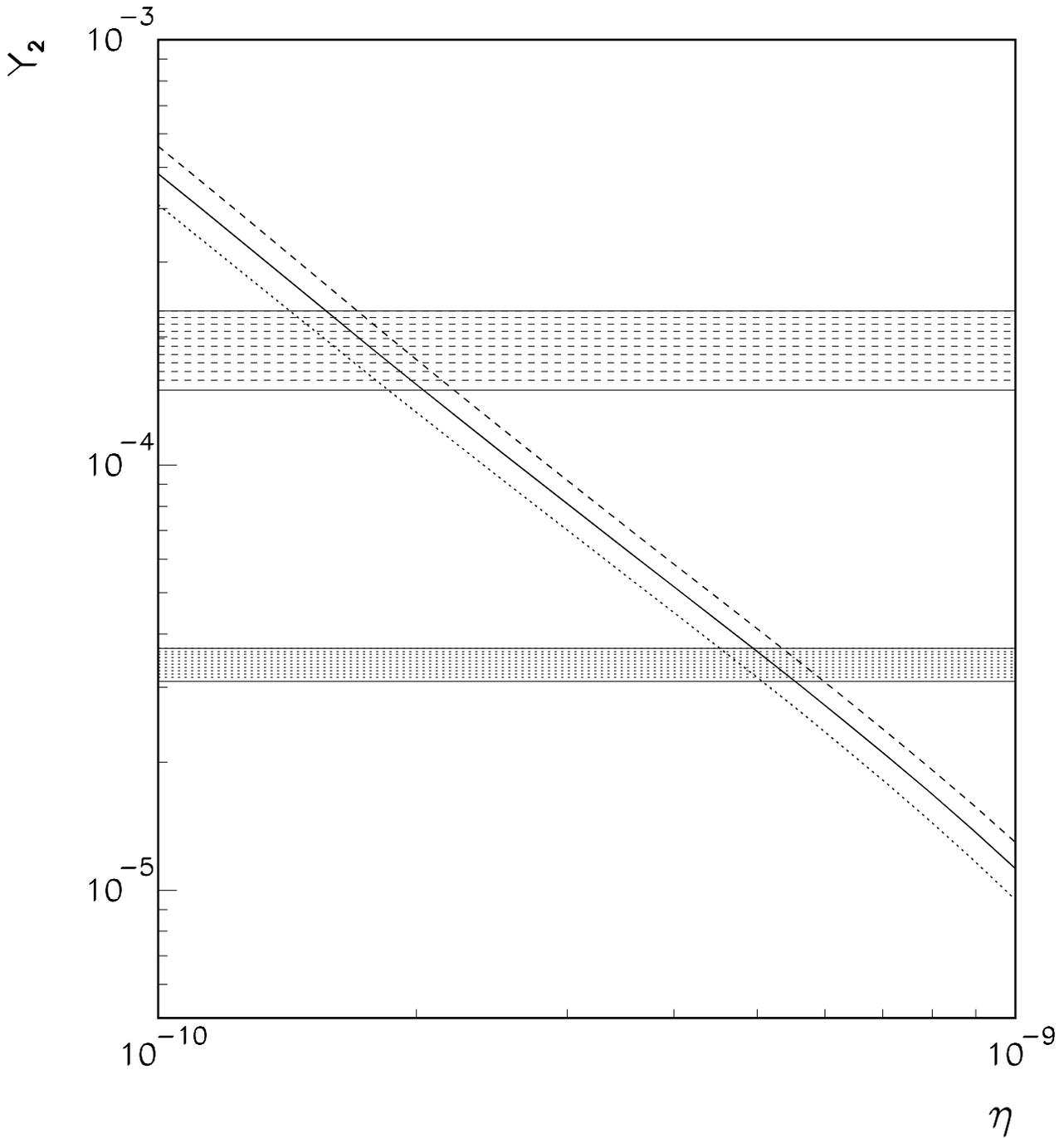} \vspace{-1.5cm}
\caption{The quantity $Y_2$ versus $\eta$ is reported. The notation used is
the same of Fig. 8. Due to the negligible dependence of $Y_2$ on small
variations of $\tau_n^{ex}$ no splitting of lines is present. The
horizontal bands dashed and dotted are the experimental values of Ref.s
\protect\cite{Songaila,Tytler}.}
\end{figure}
\begin{figure}
\epsffile{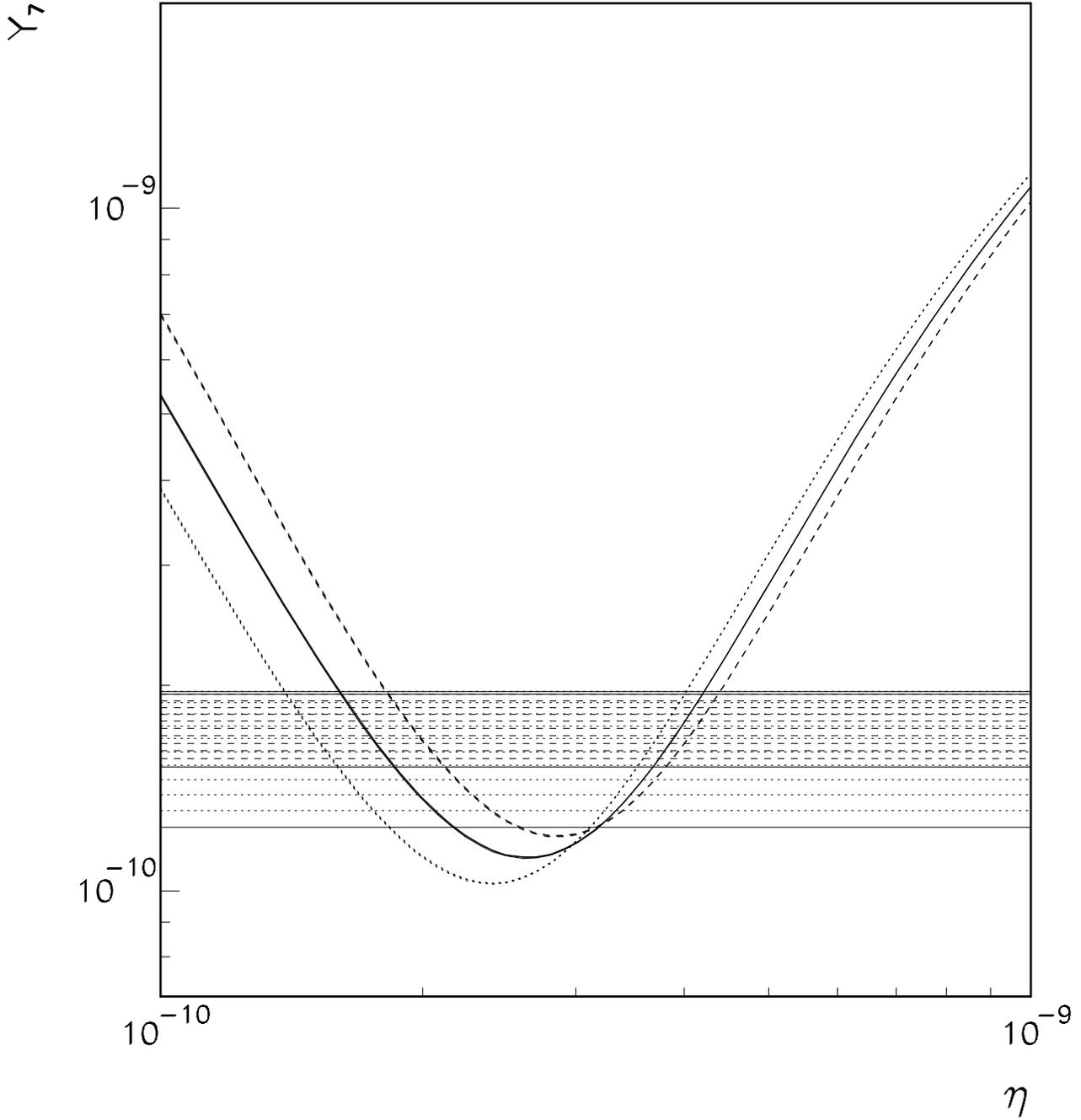} \vspace{-1.5cm}
\caption{The quantity $Y_7$ versus $\eta$. The notation used is
the same of Fig. 8. 
There is no splitting of lines related to $\Delta \tau_n$,
due to the negligible dependence of $Y_7$ on small
variations of $\tau_n^{ex}$.
The
horizontal bands dashed and dotted are the experimental values of Ref.s
\protect\cite{Thorburn, Molaro} and \protect\cite{Bonifacio}, respectively.}
\end{figure}
\begin{figure}
\epsffile{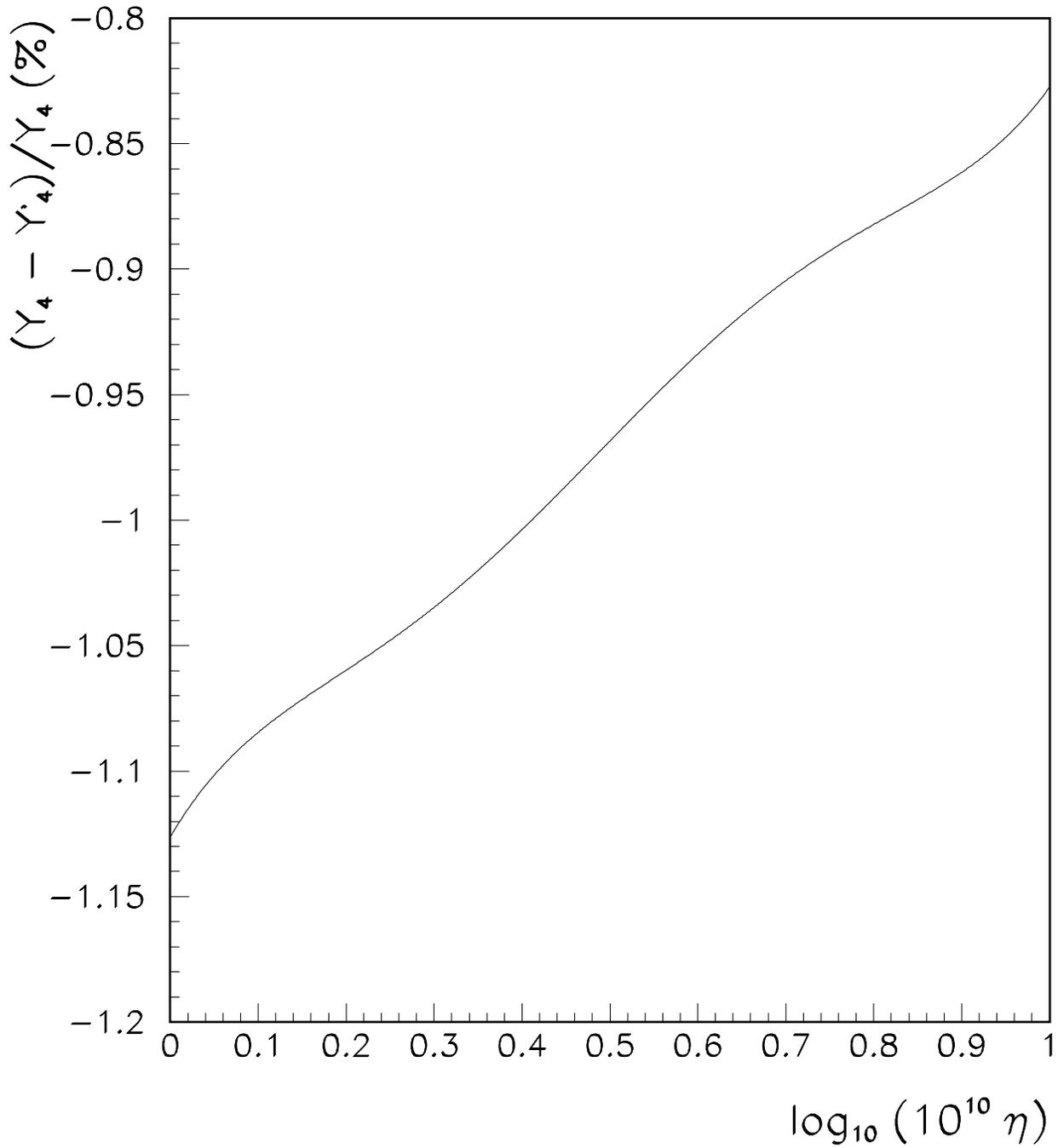}
\caption{The ratio $(Y_4-Y_4')/Y_4$ versus $\log_{10}(10^{10} \eta)$ 
for $N_\nu=3$ and
$\tau_n=885.3~s$ \protect\cite{Sarkar} (see Section
\ref{s:numres}).}
\end{figure}
\begin{figure}
\begin{tabular}{cc}
\epsfxsize=7cm
\epsfysize=7cm
\epsffile{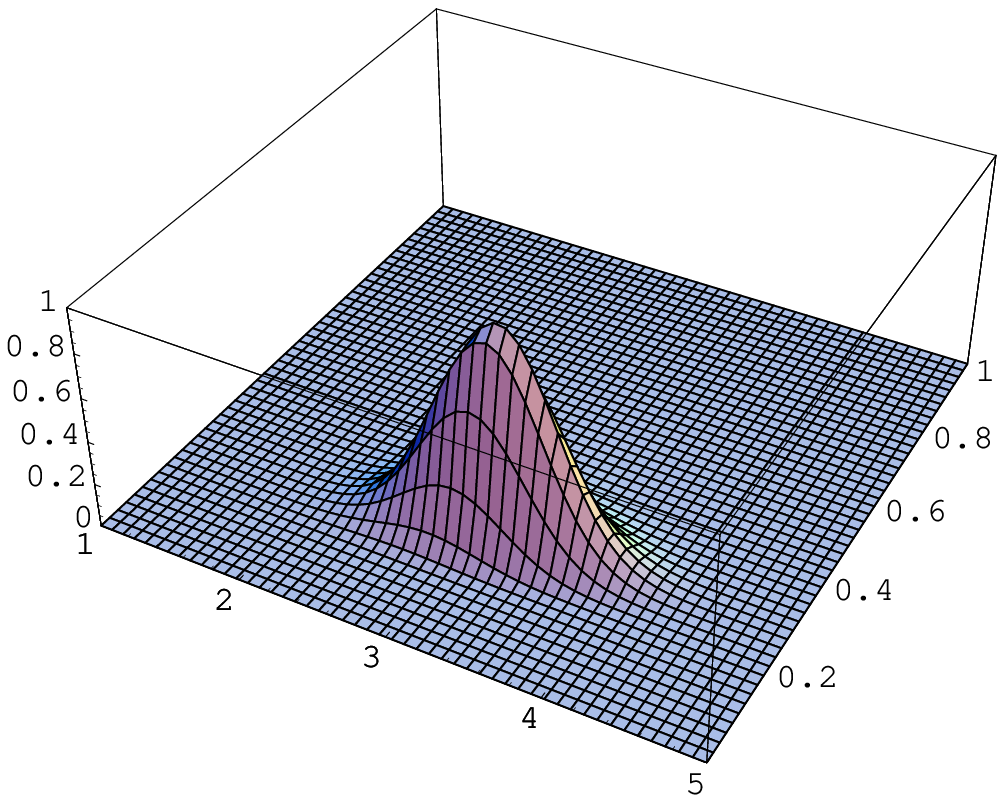} &
\epsfxsize=7cm
\epsfysize=7cm
\epsffile{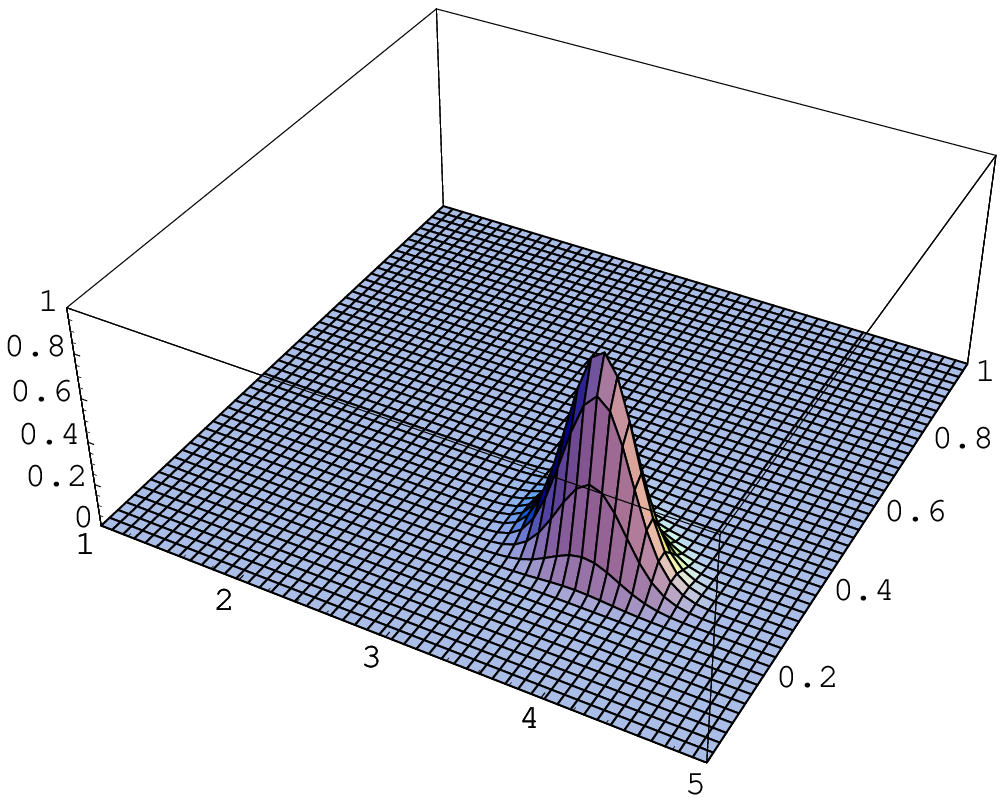} \\
\epsfxsize=7cm
\epsfysize=7cm
\epsffile{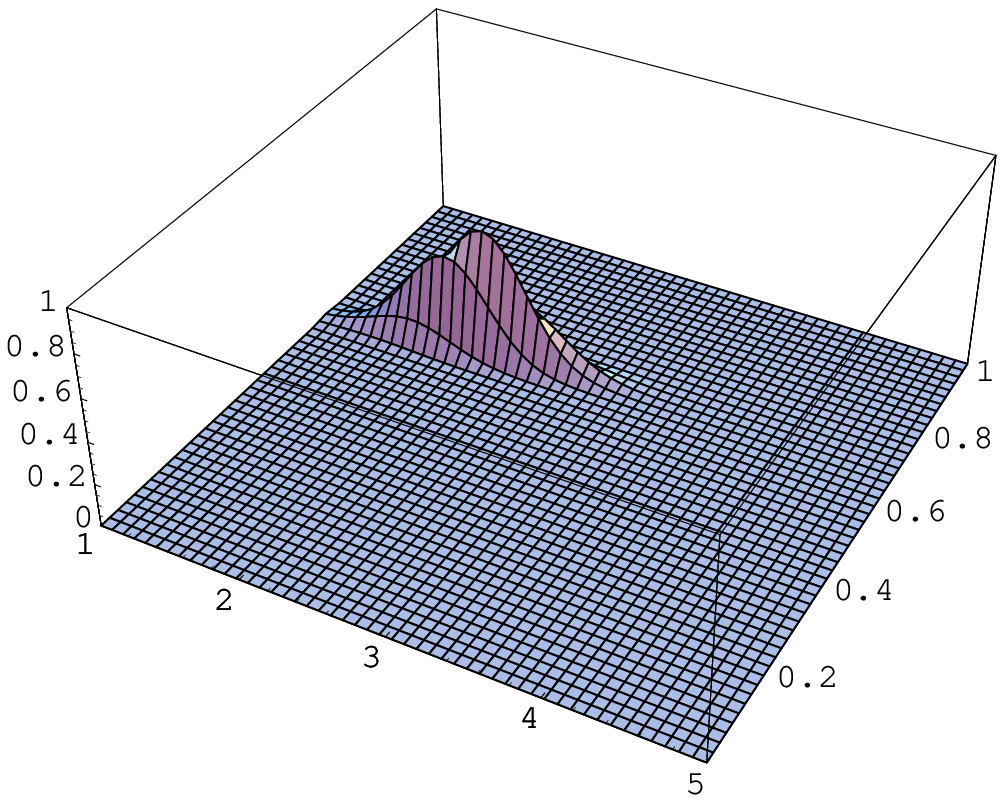} &
\epsfxsize=7cm
\epsfysize=7cm
\epsffile{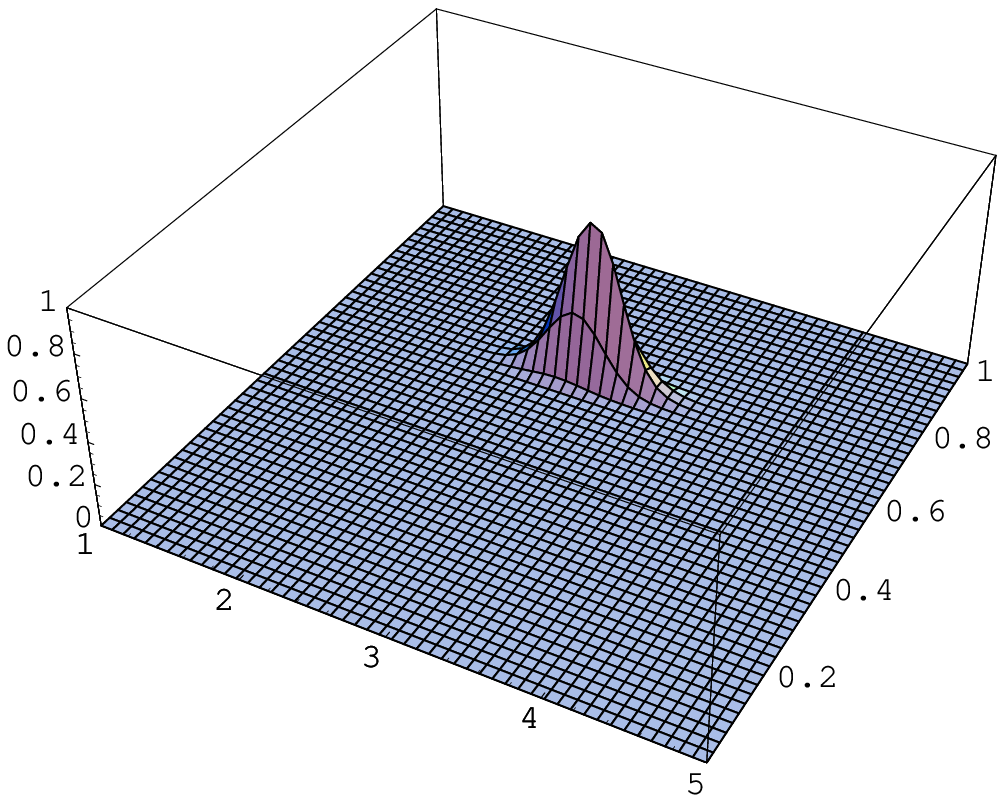}
\end{tabular}
\caption{The likelihood distributions for the light element yields $Y_2$,
$Y_4$, $Y_7$ are shown as functions of $N_\nu$ and $\log_{10}(10^{10})
\eta$, normalized to unity in correspondence of the experimental values.
From left to right and from top to bottom the
following cases are considered: a) high $D$, low $^4He$; b) high $D$, high
$^4He$; c) low $D$, low $^4He$; d) low $D$, high $^4He$. The plots for cases
c) and d) are rescaled by a factor 25 and 100 times, respectively, compared to
the one of a) and b).}
\end{figure}

\begin{thebibliography} \\

\bibitem{Wagoner} R.V. Wagoner, W.A. Fowler, and F. Hoyle, {\ApJ} {\bf 148}
(1967) 3; 

R.V. Wagoner, {\ApJS} {\bf 18} (1969) 247; R.V. Wagoner, {\ApJ}
{\bf 179} (1973) 343.

\bibitem{Steigman} B.E.J. Pagel, E.A. Simonson, R.J. Terlevich, and M.
Edmunds, {\it MNRAS} {\bf 255} (1992) 325; 

E. Skillman and R.C. Kennicutt, \ApJ {\bf 411} (1993) 655;  

E. Skillman, R.J. Terlevich, R.C. Kennicutt,
D.R. Garnett, and E. Terlevich, \ApJ {\bf 431} (1994) 172.

\bibitem{Izotov} Y.I. Izotov, T.X. Thuan, and V.A. Lipovetsky, \ApJ {\bf
435} (1994) 647; {\it Ap.J.S.} {\bf 108} (1997) 1.

\bibitem{Songaila} R.F. Carswell, M. Rauch, R.J. Weymann, A.J. Cooke, and
J.K. Webb, {\it MNRAS} {\bf 268} (1994) L1;\\ A. Songaila, L.L. Cowie, C.
Hogan, and M. Rugers, \Nature {\bf 368} (1994) 599;\\ M. Rugers and C.J.
Hogan, {\it A.J.} {\bf 111} (1996) 2135;\\  R.F. Carswell, {\it et al.} {\it
MNRAS} {\bf 278} (1996) 518;\\ E.J. Wampler, {\it et al.}, {\it A.A.} {\bf 316}
(1996) 33;\\ J.K. Webb, R.F. Carswell, K.M. Lanzetta, R. Ferlet, M.
Lemoine, A. Vidal-Madjar, and D.V. Bowen, \Nature {\bf 388} (1997) 250;\\ D.
Tytler {\it et al.}, astro-ph/9810217 (1998).

\bibitem{Tytler} D. Tytler, X.-M. Fan, and S. Burles, \Nature {\bf
381} (1996) 207;\\ S. Burles and D. Tytler, \ApJ {\bf 460} (1996) 584.

\bibitem{Thorburn} J.A. Thorburn, \ApJ {\bf 421} (1994) 318.

\bibitem{Molaro} P. Molaro, F. Primas, and P. Bonifacio, {\it Astron.
Astrophys.} {\bf 295} (1995) L47.

\bibitem{Ryan} S.G. Ryan, J.E. Norris, and T.C. Beers, \ApJ {\bf 506}
(1998) 892.

\bibitem{Bonifacio} P. Bonifacio and P. Molaro, {\it Mon. Not. Roy.
Astron. Soc.} {\bf 285} (1997) 847.

\bibitem{Sarkar} S. Sarkar, astro-ph/9903183.

\bibitem{EMMP1} S. Esposito, G. Mangano, G. Miele, and O. Pisanti, \NP
{\bf B540} (1999) 3.

\bibitem{Kawano} L. Kawano, preprint FERMILAB-Pub-88/34-A; preprint
FERMILAB-Pub-92/04-A.

\bibitem{Montecarlo} L.M. Krauss and P. Romanelli, \ApJ {\bf 358} (1990)
47;\\ P.J. Kernan and L.M. Krauss, \PRL {\bf 72} (1994) 3309.

\bibitem{Fiorentini} G. Fiorentini, E. Lisi, S. Sarkar, and F.L. Villante,
\PR {\bf D58} (1998) 063506.

\bibitem{Lopez} R.E. Lopez and M.S. Turner, \PR {\bf D59} (1999) 103502.

\bibitem{PDG98}  C. Caso \etal, {\it Eur. Phys. Jour.} {\bf C3} (1998) 1.

\bibitem{FTQFT} D.A. Dicus,  E.W. Kolb , A.M. Gleeson, E.C.G. Sudarshan,
V.L. Teplitz, and M.S. Turner,  \PR {\bf D26} (1982) 2694;\\ J.L. Cambier,
J.R. Primack, and M. Sher, \NP {\bf B209} (1982) 372;\\ J.F. Donoghue, B.R.
Holstein, and R.W. Robinett, {\it Ann. Phys.} (N.Y.) {\bf 164} (1985) 23;\\
J.F. Donoghue and B.R. Holstein, \PR {\bf D28} (1983) 340 and \PR {\bf D29}
(1984) 3004;\\ A.E. Johansson, G. Peresutti, and B.S. Skagerstam, \NP {\bf
B278} (1986) 324;\\ W. Keil, \PR {\bf D40} (1989) 1176;\\ R. Baier, E.
Pilon, B. Pire, and D. Schiff, {\it Nucl. Phys.} {\bf B 336} (1990) 157;\\
W. Keil and R. Kobes, {\it Physica} {\bf A 158} (1989) 47;\\ M. LeBellac
and D. Poizat, {\it Z. Phys.} {\bf C47} (1990) 125;\\ T. Altherr and
P.Aurenche,
\PR {\bf D40} (1989) 4171;\\ R.L. Kobes and G.W. Semeneff, \NP {\bf B260}
(1985) 714 and {\bf B272} (1986) 329;\\ R.F. Sawyer, \PR {\bf D53} (1996)
4232;\\ I.A. Chapman, \PR {\bf D55} (1997) 6287;\\ S. Esposito, G. Mangano,
G. Miele, and O. Pisanti, \PR {\bf D58} (1998) 105023.

\bibitem{Sirlin} A. Sirlin, \PR {\bf 164} (1967) 1767.

\bibitem{Marciano} W.J. Marciano and A. Sirlin, \PRL {\bf
56} (1986) 22.

\bibitem{Marciano2} W.J. Marciano and A. Sirlin, \PRL
{\bf 46} (1981) 163.

\bibitem{Wilkinson} D.M. Wilkinson, \NP {\bf A377} (1982)
474.

\bibitem{Seckel} D. Seckel, preprint BA-93-16, hep-ph/9305311;\\ R. E.
Lopez, M. S. Turner, and G. Gyuk, \PR {\bf D56} (1997) 3191.

\bibitem{Enqvist} K. Enqvist, K. Kainulainen, and V. Semikoz,
\NP {\bf B374} (1992) 392.

\bibitem{numrec} B.P. Flannery, W.H. Press, S.A. Teukolsky, and W.T.
Vetterling, {\it Numerical Recipes in Fortran}, Cambridge University Press.

\bibitem{network} For the expressions of the
rates see the Web sites:\\ 
{\it http://www.phy.ornl.gov/astrophysics/data/data.html} or \\
{\it http://pntpm.ulb.ac.be/Nacre/barre\_database.htm}.

\bibitem{neutrdec}M.A. Herrera and S. Hacyan, {\it Ap. J.} {\bf 336}
(1989) 539;\\
 N.C. Rana and B. Mitra, {\it Phys. Rev.} {\bf D44} (1991) 393;\\
 S. Dodelson and M.S. Turner, {\it Phys. Rev.} {\bf D46} (1992) 3372; \\
 A.D. Dolgov and M. Fukugita, {\it Phys. Rev.} {\bf D46} (1992) 5378; \\
 N.Y. Guedin and O.Y. Guedin, astro-ph/9712199;\\
 S. Hannestad and J. Madsen, {\it Phys. Rev.} {\bf D52} (1995) 1764; \\
 A.D. Dolgov, S.H. Hansen, and D.V. Semikoz, {\it Nucl. Phys.} {\bf B503}
 (1997) 426.

\bibitem{Fields}
 B. Fields, S. Dodelson, and M.S. Turner, {\it Phys. Rev.} {\bf D47} (1993)
 4309.

\bibitem{photon} K. Ahmed and S.S. Masood, {\it Annals Phys.} {\bf 207}
(1991) 460.

\end{thebibliography}
\end{document}